\input epsf.tex

\catcode`@=11
%
%---------------------------- \lsim \gsim ----------------------------------
%
%    Simboli di minore o circa uguale, maggiore o circa uguale.
%
\def\lsim{\mathchoice
  {\mathrel{\lower.8ex\hbox{$\displaystyle\buildrel<\over\sim$}}}
  {\mathrel{\lower.8ex\hbox{$\textstyle\buildrel<\over\sim$}}}
  {\mathrel{\lower.8ex\hbox{$\scriptstyle\buildrel<\over\sim$}}}
  {\mathrel{\lower.8ex\hbox{$\scriptscriptstyle\buildrel<\over\sim$}}} }
\def\gsim{\mathchoice
  {\mathrel{\lower.8ex\hbox{$\displaystyle\buildrel>\over\sim$}}}
  {\mathrel{\lower.8ex\hbox{$\textstyle\buildrel>\over\sim$}}}
  {\mathrel{\lower.8ex\hbox{$\scriptstyle\buildrel>\over\sim$}}}
  {\mathrel{\lower.8ex\hbox{$\scriptscriptstyle\buildrel>\over\sim$}}} }
\def\croce{\displaystyle / \kern-0.2truecm\hbox{$\backslash$}}
\def\lqua{\lower4pt\hbox{\kern5pt\hbox{$\sim$}}\raise1pt
\hbox{\kern-8pt\hbox{$<$}}~}
\def\gqua{\lower4pt\hbox{\kern5pt\hbox{$\sim$}}\raise1pt
\hbox{\kern-8pt\hbox{$>$}}~}
\def\mma{\lower1pt\hbox{\kern5pt\hbox{$\scriptstyle <$}}\raise2pt
\hbox{\kern-7pt\hbox{$\scriptstyle >$}}~}
\def\mmb{\lower1pt\hbox{\kern5pt\hbox{$\scriptstyle >$}}\raise2pt
\hbox{\kern-7pt\hbox{$\scriptstyle <$}}~}
\def\mmc{\lower4pt\hbox{\kern5pt\hbox{$<$}}\raise1pt
\hbox{\kern-8pt\hbox{$>$}}~}
\def\mmd{\lower4pt\hbox{\kern5pt\hbox{$>$}}\raise1pt
\hbox{\kern-8pt\hbox{$<$}}~}
\def\lsu{\raise4pt\hbox{\kern5pt\hbox{$\sim$}}\lower1pt
\hbox{\kern-8pt\hbox{$<$}}~}
\def\gsu{\raise4pt\hbox{\kern5pt\hbox{$\sim$}}\lower1pt
\hbox{\kern-8pt\hbox{$>$}}~}
\def\croce{\displaystyle / \kern-0.2truecm\hbox{$\backslash$}}
\def\ali{\hbox{A \kern-.9em\raise1.7ex\hbox{$\scriptstyle \circ$}}}
\def\2frecce{\hbox{\lower 0.3ex\hbox{$\leftarrow$} 
\hbox{\kern-1.3em\raise 0.3ex\hbox{$\rightarrow$}}}}
%
%
%---------------------------  \quadratello ---------------------------------
%
%
\def\quad@rato#1#2{{\vcenter{\vbox{
        \hrule height#2pt
        \hbox{\vrule width#2pt height#1pt \kern#1pt \vrule width#2pt}
        \hrule height#2pt} }}}
\def\quadratello{\mathchoice
\quad@rato5{.5}\quad@rato5{.5}\quad@rato{3.5}{.35}\quad@rato{2.5}{.25} }
%
%------------------------ caratteri grassetto speciali -----------
%
\font\s@=cmss10\font\s@b=cmbx8
\def\reali{{\hbox{\s@ l\kern-.5mm R}}}
\def\m{{\hbox{\s@ l\kern-.5mm M}}}
\def\k{{\hbox{\s@ l\kern-.5mm K}}}
\def\naturali{{\hbox{\s@ l\kern-.5mm N}}}
\def\interi{{\mathchoice
 {\hbox{\s@ Z\kern-1.5mm Z}}
 {\hbox{\s@ Z\kern-1.5mm Z}}
 {\hbox{{\s@b Z\kern-1.2mm Z}}}
 {\hbox{{\s@b Z\kern-1.2mm Z}}}  }}
\def\complessi{{\hbox{\s@ C\kern-1.7mm\raise.4mm\hbox{\s@b l}\kern.8mm}}}
\def\toro{{\hbox{\s@ T\kern-1.9mm T}}}
\def\unity{{\hbox{\s@ 1\kern-.8mm l}}}
%
%------------------------- bold math. it. --------------------------
%
\font\bold@mit=cmmi10
\def\setbmit{\textfont1=\bold@mit}
\def\bmit#1{\hbox{\textfont1=\bold@mit$#1$}}
%
%-----------------------------------------------------------------
\catcode`@=12

\magnification=1200
\hsize=15truecm
\vsize=23truecm
\baselineskip 18 truept
\voffset=-0.5truecm
\parindent=0cm
\overfullrule=0pt

\def\Ai{\hbox{\hbox{${\cal A}$}}\kern-1.9mm{\hbox{${/}$}}}
\def\Vi{\hbox{\hbox{${\cal V}$}}\kern-1.9mm{\hbox{${/}$}}}
\def\Di{\hbox{\hbox{${\cal D}$}}\kern-1.9mm{\hbox{${/}$}}}
\def\lam{\hbox{\hbox{${\lambda}$}}\kern-1.6mm{\hbox{${/}$}}}
\def\D{\hbox{\hbox{${D}$}}\kern-1.9mm{\hbox{${/}$}}}
\def\A{\hbox{\hbox{${A}$}}\kern-1.8mm{\hbox{${/}$}}}
\def\V{\hbox{\hbox{${V}$}}\kern-1.9mm{\hbox{${/}$}}}
\def\parz{\hbox{\hbox{${\partial}$}}\kern-1.7mm{\hbox{${/}$}}}
\def\B{\hbox{\hbox{${B}$}}\kern-1.7mm{\hbox{${/}$}}}
\def\R{\hbox{\hbox{${R}$}}\kern-1.7mm{\hbox{${/}$}}}
\def\si{\hbox{\hbox{${\xi}$}}\kern-1.7mm{\hbox{${/}$}}}
\def\Oi{\hbox{\hbox{${\cal O}$}}\kern-1.9mm{\hbox{${/}$}}}

\catcode`@=11
%
%---------------------------- \lsim \gsim ----------------------------------
%
%    Simboli di minore o circa uguale, maggiore o circa uguale.
%
\def\lsim{\mathchoice
  {\mathrel{\lower.8ex\hbox{$\displaystyle\buildrel<\over\sim$}}}
  {\mathrel{\lower.8ex\hbox{$\textstyle\buildrel<\over\sim$}}}
  {\mathrel{\lower.8ex\hbox{$\scriptstyle\buildrel<\over\sim$}}}
  {\mathrel{\lower.8ex\hbox{$\scriptscriptstyle\buildrel<\over\sim$}}} }
\def\gsim{\mathchoice
  {\mathrel{\lower.8ex\hbox{$\displaystyle\buildrel>\over\sim$}}}
  {\mathrel{\lower.8ex\hbox{$\textstyle\buildrel>\over\sim$}}}
  {\mathrel{\lower.8ex\hbox{$\scriptstyle\buildrel>\over\sim$}}}
 {\mathrel{\lower.8ex\hbox{$\scriptscriptstyle\buildrel>\over\sim$}}} }

\def\gsu{\raise4pt\hbox{\kern5pt\hbox{$\sim$}}\lower1pt\hbox{\kern-8pt
\hbox{$>$}}~}
\def\lsu{\raise4pt\hbox{\kern5pt\hbox{$\sim$}}\lower1pt\hbox{\kern-8pt
\hbox{$<$}}~}

\def\croce{\displaystyle / \kern-0.2truecm\hbox{$\backslash$}}
\def\lqua{\lower4pt\hbox{\kern5pt\hbox{$\sim$}}\raise1pt
\hbox{\kern-8pt\hbox{$<$}}~}
\def\gqua{\lower4pt\hbox{\kern5pt\hbox{$\sim$}}\raise1pt
\hbox{\kern-8pt\hbox{$>$}}~}
\def\mma{\lower1pt\hbox{\kern5pt\hbox{$\scriptstyle <$}}\raise2pt
\hbox{\kern-7pt\hbox{$\scriptstyle >$}}~}
\def\mmb{\lower1pt\hbox{\kern5pt\hbox{$\scriptstyle >$}}\raise2pt
\hbox{\kern-7pt\hbox{$\scriptstyle <$}}~}
\def\mmc{\lower4pt\hbox{\kern5pt\hbox{$<$}}\raise1pt
\hbox{\kern-8pt\hbox{$>$}}~}
\def\mmd{\lower4pt\hbox{\kern5pt\hbox{$>$}}\raise1pt
\hbox{\kern-8pt\hbox{$<$}}~}
\def\croce{\displaystyle / \kern-0.2truecm\hbox{$\backslash$}}
%
%
%---------------------------  \quadratello ---------------------------------
%
%
\def\quad@rato#1#2{{\vcenter{\vbox{
        \hrule height#2pt
        \hbox{\vrule width#2pt height#1pt \kern#1pt \vrule width#2pt}
        \hrule height#2pt} }}}
\def\quadratello{\mathchoice
\quad@rato5{.5}\quad@rato5{.5}\quad@rato{3.5}{.35}\quad@rato{2.5}{.25} }
%
%------------------------ caratteri grassetto speciali -----------
%
\font\s@=cmss10\font\s@b=cmbx8
\def\reali{{\hbox{\s@ l\kern-.5mm R}}}
\def\m{{\hbox{\s@ l\kern-.5mm M}}}
\def\k{{\hbox{\s@ l\kern-.5mm K}}}
\def\naturali{{\hbox{\s@ l\kern-.5mm N}}}
\def\interi{{\mathchoice
 {\hbox{\s@ Z\kern-1.5mm Z}}
 {\hbox{\s@ Z\kern-1.5mm Z}}
 {\hbox{{\s@b Z\kern-1.2mm Z}}}
 {\hbox{{\s@b Z\kern-1.2mm Z}}}  }}
\def\complessi{{\hbox{\s@ C\kern-1.7mm\raise.6mm\hbox{\s@b l}\kern.8mm}}}
\def\toro{{\hbox{\s@ T\kern-1.9mm T}}}
\def\unity{{\hbox{\s@ 1\kern-.8mm l}}}
%
%------------------------- bold math. it. --------------------------
%
\font\bold@mit=cmmib10
\def\setbmit{\textfont1=\bold@mit}
\def\bmit#1{\hbox{\textfont1=\bold@mit$#1$}}
%
%-----------------------------------------------------------------
\catcode`@=12

\null
\vskip 0.5truecm

\centerline{\bf GAUGE--INVARIANT CHARGED, MONOPOLE AND} 

\centerline{\bf DYON FIELDS IN GAUGE THEORIES}
\vskip 1truecm
\centerline{J. FR\"OHLICH}

\vskip 0.3truecm
\centerline{\sl Theoretical Physics, ETH--H\"onggerberg, CH--8093 Z\"urich,
Switzerland}

\vskip 0.5truecm
\centerline{P.A. MARCHETTI}
\vskip 0.3truecm
\centerline{\sl Dipartimento di Fisica, Universit\`a  di Padova}

\centerline{\sl and}

\centerline{\sl INFN -- Sezione di Padova, I--35131 Padova, Italy}
\vskip 2truecm

\centerline{\bf Abstract}
\vskip 0.3truecm
We propose explicit recipes to construct the euclidean Green functions 
of gauge--invariant charged, monopole and dyon fields in 
four--dimensional gauge theories whose phase diagram contains phases with 
deconfined electric and/or magnetic charges. In theories with only either 
abelian electric or magnetic charges, our construction is an 
euclidean version of Dirac's original proposal, the magnetic dual of 
his proposal, respectively. Rigorous mathematical control is achieved for 
a class of abelian lattice theories. In theories where electric and
magnetic charges
coexist, our construction of Green functions of electrically or
magnetically charged fields involves taking an average over 
Mandelstam strings or the dual magnetic flux tubes, in accordance with 
Dirac's flux quantization condition. We apply our construction to 't 
Hooft--Polyakov monopoles and Julia--Zee dyons. Connections between our
construction and the 
semiclassical approach are discussed.

\vfill\eject

\vskip 0.3truecm
{\bf 1.Introduction}
\vskip 0.3truecm

In this paper, we study a variety of gauge theories in four dimensions with
the following features: their phase diagrams contain phases with deconfined
electric or magnetic charges; their particle spectra thus contain 
electrically or magnetically charged particles or dyons. They may exhibit 
phase transitions characterized by condensation of charged particles or 
magnetic monopoles. Some of them exhibit duality symmetries.

One is interested in studying various aspects of the phase diagram and of 
the dynamics in particular phases of such theories. Obviously, one 
would like to study these theories analytically. However, explicit 
analytical results are often only available for physically rather 
unrealistic theories with supersymmetries. In order to study more realistic
theories, one therefore often resorts to numerical investigations of 
lattice approximations of such theories. Usually, such investigations are 
based on a euclidean (imaginary--time) formulation of quantum field theory
obtained from a real--time formulation by Wick rotation. In the lattice 
approximation, one replaces Euclidean space--time by a (finite, but 
arbitrarily large) lattice.

In order to study the mass spectrum of a quantum field theory, one 
considers euclidean Green functions of gauge--invariant (physical) fields 
of the theory that couple a vacuum (ground) state to some one--particle 
state. Masses can be calculated from exponential decay rates of certain 
two--point euclidean Green functions.

Signals for a phase transition, e.g., one between a Coulomb-- and a 
confining phase, can be detected in an analysis of asymptotic behaviour of 
suitable euclidean Green functions. For example, the transition from a 
Coulomb-- to a confining phase in an abelian gauge theory is reflected in 
the appearance of long--range order in the two--point euclidean Green 
functions of magnetic monopoles.

On a more foundational level, we are longing for a complete description of
quantum field theories in terms of the (euclidean) Green functions of 
gauge--invariant interpolating fields. For example, to describe the deconfined 
phase of an abelian gauge theory, we would like to construct Green 
functions of charged fields. 

For QED--like theories without dynamical magnetic monopoles, a proposal 
for a gauge--invariant charged field has been made, many years ago, by 
Dirac [1]. For lattice theories of this type, Dirac's proposal has been studied
carefully e.g. in [2,3]. For the convenience of the reader, the main 
results of this analysis are summarized in sect. 3.

There are, however, plenty of physically interesting theories with 
\underbar{deconfined}, electrically charged particles \underbar{and} dynamical
magnetic monopoles. The problem of constructing gauge--invariant electrically 
or magnetically charged interpolating fields for such theories has not
been solved in adequate generality and is quite non--trivial. The 
difficulties encountered in studying this problem are a consequence of the 
Dirac quantization condition. We report partial results towards a solution 
of this problem in sect. 4. 

The main purpose  of this paper is to describe constructions of 
gauge--invariant electrically or magnetically charged interpolating fields 
and of dyon fields in (lattice) gauge  theories with dynamical electric 
charges and dynamical magnetic monopoles and to exhibit duality 
transformations converting electrically into magnetically charged fields 
(and vice versa). Our work is primarily \underbar{kinematical}: we propose 
fairly explicit recipes for how to construct the 
euclidean Green functions of such fields. But we do not engage in any 
mathematically careful analysis of the properties of these Green functions.
Instead, we gather evidence supporting various conjectures on their 
behaviour. 
Part of this evidence  comes from previous, mathematically precise work on 
lattice gauge  theory, another part is based on more heuristic arguments 
and conventional wisdom.

All in all, we believe we arrive at a fairly consistent picture of how 
electrically or magnetically charged and dyon fields should be constructed.

We also review duality properties of some gauge theories and analyze how 
duality transformations act on charged and dyon fields. This part of our 
analysis makes contact with issues that have been quite topical, during the
past few years; (see e.g.[4]).

Finally, we outline a formal construction of Green functions of monopole--
and dyon fields in the continuum $SU(2)$ Georgi--Glashow model and indicate 
how our construction is related to the (semi--)classical  analyses of 't 
Hooft and Polyakov [5] and of Julia and Zee [6].

Next, we present brief summaries of the different sections of this paper. 

In section 2, we introduce the gauge theories studied in this paper. We 
consider three classes of models (A, B, and C). The models in class A are 
\underbar{non--compact} abelian lattice gauge theories with electrically charged
matter fields, but without dynamical magnetic monopoles, i.e., models of 
lattice QED. The models in class B are \underbar{compact} abelian lattice gauge 
theories with electrically charged matter fields. As originally pointed out
by Polyakov [7], they describe dynamical magnetic monopoles coexisting with
electrically charged particles.
The models in class C are related to (lattice approximations or formal 
continuum limits of) the Georgi--Glashow model.
We define our models in terms of euclidean action functionals. We 
introduce some key notation and review some facts on the phase diagrams of 
these models.

In section 3, we construct gauge--invariant charged fields for models of 
class A, in accordance with Dirac's proposal and with the result of 
[2,3]. We also construct monopole fields in compact, pure abelian lattice gauge 
theories. These theories are dual, in the sense of Kramers--Wannier 
duality [8], to some models of class A, and this suggests a dual variant of 
Dirac's proposal as the right definition of monopole fields.

We start section 3 with a short recapitolation of Osterwalder--Schrader 
reconstruction and of an analysis of superselection sectors based on 
euclidean Green functions of charged fields and monopole fields. We then 
recall (an euclidean version of) Dirac's proposal [1] for gauge--invariant 
charged fields, 

$$
\Phi (x, E) = \Phi_x e^{i(A,E(x))} \eqno(1.1)
$$

where $\Phi_x$ is a charged matter field, and $A$ is a (non--compact, i.e.,
real--valued) abelian gauge potential; furthermore, $E(x)$ is a c--number 
one--form related to the electrostatic Coulomb field of a point charge. 
(Thus, $E(x)$ has a source at the point $x$ whose charge is the same as the
charge of $\Phi$). We study the ``infra--particle nature" of charged 
particles in such theories.

Our construction of monopole fields and our analysys of their Green 
functions in compact abelian gauge theories without matter fields follows 
from the results on electrically charged field by using Kramers--Wannier 
duality; (this is made explicit in sect. 3.3).

In sect. 4, we study electrically charged particles and dynamical magnetic 
monopoles in compact abelian lattice gauge theories \underbar{with} matter 
fields. We start by explaining the origin of Dirac's quantization 
condition,

$$
q_e \cdot q_m = 2\pi n
\eqno(1.2)
$$

$n= 0, \pm 1, \pm 2, ...$. We  then construct gauge--invariant electrically
charged fields as averages of charged fields multiplied by Mandelstam 
string operators (exponential of the gauge field integrated along a path, 
called Mandelstam string [9]), the average being taken over a suitable space of
Mandelstam strings. Among the more subtle points appearing in this paper is
the one to come up with a good definition of  ``averages over Mandelstam 
strings". 
In an analysis of Green functions of these fields based on successively 
integrating out the large--frequency (short--distance) modes of the 
fields, one observes that the large--distance effective theory becomes 
increasingly similar to the one describing a model in class A. In 
particular, the functional integral expressions for these Green functions 
resemble the ones for the Green functions of charged fields constructed 
according to Dirac's proposal, in models of class A. Suitable averaging 
over Mandelstam strings yields operators which, under renormalization, approach 
ones related to (1.1). The magnetic monopoles get suppressed, and Dirac's 
flux quantization condition becomes irrelevant. 

We then proceed to define monopole fields, or, more precisely, euclidean 
Green functions of such fields, in terms of certain averages over 't Hooft 
disorder operators [10]. We also present a heuristic analysis of properties of 
euclidean Green functions of electrically charged-- and monopole fields 
and compare our results with these in section 3.

We conclude section 4 with an analysis of dyons and of an $SL(2, {\bf Z})$ 
duality group in a compact abelian lattice gauge theory with a topological 
term, related to the instanton number, in the action. Our analysis is based
on previous work of Cardy and Rabinovici [11].

In section 5, we outline a construction of physical interpolating fields 
for 't Hooft--Polyakov monopoles and Julia--Zee dyons in the 
Georgi--Glashow model on the lattice and in the formal continuum limit of 
this model. The role of 't Hooft's ${\bf Z}_2$ monopoles and the 
corresponding disorder operators in our construction in explained in 
detail.

Some technical points are relegated to two short appendices.

\vskip 0.3truecm
{\bf 2. The models}
\vskip 0.3truecm

The models we consider in this paper are lattice gauge theories.( Some 
comments on a continuum gauge theory are added in the last section, for the 
Georgi--Glashow model.)

Our (euclidean space--time) lattice is ${\bf Z}^4_{1\over2}$,
where the subscript ${1\over 2}$ indicates that 
the coordinates of the sites are half--integers. We will also have to 
consider sublattices (more precisely cell subcomplexes) of
${\bf Z}^4_{1\over2}$. 

Lattice fields can be defined as follows:

A scalar field $\Phi$ is a map from the sites, $i$, 
of the lattice to a normed vector space $V_H$.

A fermion field $\Psi$ is an anticommuting map from the sites to the 
orthonormal frames of a vector space $V_s \otimes V_F$, the fermion space, 
where $V_s$, the spin space, carries a representation of the 
Dirac--Clifford algebra.

A gauge field $g$ is a map from the links  $<ij>$ of the lattice to a group
$G$, the gauge group.

If $W$ is an additive abelian group and $k$ is a positive integer, one can 
define $k-$ forms with values in $W$ as maps, $F_k$, from oriented
$k$-dimensional cells, $c_k$ of the lattice to $W$
satisfying $F(-c_k)=-F(c_k)$, where 
$c_k$ denotes the cell obtained from $c_k$ by reversing the orientation.

We denote by $d$ the lattice exterior differential

$$
dF(c_{k+1})=\sum_{c_k \in \partial c_{k+1}} F(c_k)
$$

and by $*$ the 
Hodge--star:

Let $c^*_{4-k}$ denote the cell in the dual lattice, ${\bf 
Z}^4$, dual to $c_k$. Then

$$
^*F(c^*_{4-k})=F(c_k)
$$

We also introduce the codifferential $\delta=(-)^{d(k+1)}{^*d^*}$ and 
the Laplacian 
$\Delta=d \delta+ \delta d$. 
If $\Lambda$ is a sublattice we denote the restriction of lattice 
operators to forms over $\Lambda$ by a subindex $\Lambda$,
e.g. $d_\Lambda$ instead of $d$.  

If $W$ is a Hilbert space with scalar 
product $(\cdot , \cdot)$ one can define a scalar product among 
$k-$forms $F$ and $F^\prime$ by

$$
(F, F^\prime) = \sum_{c_k} \Bigl(F(c_k), F^\prime (c_k) \Bigr) \eqno(2.1)
$$

The restriction of the scalar product (2.1) to forms defined on a sublattice
$\Lambda$ is 
denoted by $(\cdot,\cdot)_\Lambda$.
We define the $\ell_2$--norm of $F$ by

$$
||F||={\sqrt (F,F)}
$$

The vacuum functional of a gauge theory with gauge field $g$ and
with matter fields $\Phi, 
\Psi$ is given in terms of formal integration ``measures"

$$
d\mu (g, \Phi, \bar\Psi,\Psi)= {1\over Z} \prod_{<ij>} dg_{<ij>} \prod_i
d\Phi_i d\bar\Psi_i d \Psi_i e^{- S(g,\Phi, \bar\Psi, \Psi)} \eqno(2.2)
$$

where $dg_{<ij>}$ is the Haar measure on $G$, $d\Phi_i$ is the Lebesgue 
measure on $V_H, d\bar\Psi_i d\Psi_i$ denotes Berezin integration, $S$ is 
the total action, and $Z$ is a normalization factor, the partition function. 
[Mathematically, the measure (2.2) is first defined for a finite 
lattice $\Lambda \subset {\bf Z}^4_{1\over 2}$, with suitable  
boundary conditions at 
$\partial\Lambda$; subsequently, one takes the limit $\Lambda\nearrow {\bf 
Z}^4_{1\over 2}$].

Expectation values w.r.t. the measure (2.2) are denoted by $\langle \cdot 
\rangle$.

\vskip 0.3truecm
{\sl 2.1 Actions}
\vskip 0.3truecm

We consider the following classes of models. 

\underbar{Class A: Non--compact abelian gauge theories with matter 
fields}.

As an example we discuss the abelian Higgs model .

The fields of this model are a real gauge field, $A$, and a complex 
scalar field, $\Phi$.

The action is given by

$$
S(A, \Phi) = S_0 (A) + S_1 (A, \Phi) + S_{gf} (A) 
$$
$$
S_0 (A) = {1\over 2\beta} ||dA||^2 , \qquad
S_1 (A, \Phi) = {\kappa\over 2}
\sum_{<ij>} |\Phi_i - e^{i A_{<ij>}} \Phi_j |^2 + 
\lambda \sum_i (|\Phi_i|^2 -1)^2
\eqno(2.3)
$$

$S_{gf} (A)$ is a gauge fixing term, e.g.

$$
S_{gf} (A) = {1\over 2} ||\delta A||^2
$$

From now on, this gauge fixing term will usually not be written 
explicitly in our formulas, anymore.
In the limit $\lambda\nearrow \infty$ this model reduces to the St\"uckelberg 
model. For simplicity, we only discuss the model at $\lambda =
\infty$, in this paper. It will be referred to as model A.

\vskip 0.3 truecm
\underbar{Remark 2.1}  A second interesting model in class A is spinor 
QED on the lattice; 
(see [2] for a rather detailed discussion). 
The fields of QED are the real gauge field $A$ and a four--component fermion
field $\Psi = \{\Psi_\alpha, \alpha=1, ... 4\}$. The action is given by

$$
S(A, \bar\Psi, \Psi) = S_0 (A) + S_1 (A, \bar\Psi, \Psi) + S_{gf} (A)
$$
$$
S_1 (A, \bar\Psi, \Psi) = \kappa \sum_{<ij>} \bar\Psi_i \Gamma_{<ij>} e^{i 
A_{<ij>}} \Psi_i + \sum_i \bar\Psi_i \Psi_i \eqno(2.4)
$$

where $\Gamma_{<ij>}$ are matrices on $V_F$ given by

$$
\Gamma_{<ij>} = 1 \pm \gamma_\mu
$$

if $<ij>$
is directed in the $\pm\mu$ direction, $\mu=1, ..., d$, where
$\gamma_\mu$ are Euclidean Dirac matrices.

\vskip 0.3truecm
\underbar{Remark 2.2}
In these models, we can omit gauge fixing by working directly 
with a measure defined on gauge equivalence classes 

$$
[A] = \{A^\prime: A^\prime - A = d\xi \}, \eqno(2.5)
$$

where $\xi$ is a real scalar field [12].
\vskip 0.3truecm

\underbar{Class B: Compact abelian gauge theories with matter fields}. 

As an example we analyze the compact abelian Higgs model (model B).

In the ``Villain formulation", the basic fields are a $U(1)$--valued 
$0$--form
$\varphi$, a $U(1)$-valued 1--form $\theta$, a $2\pi {\bf Z}$--valued 1--form
$\ell$ and a $2\pi {\bf Z}$-valued 2--form $n$.  The action is given by

$$
S(\theta, n, \varphi, \ell) = S_0 (\theta, n) + S_1 (\theta, \varphi, \ell)
$$
$$
S_0 (\theta, n) ={\beta\over 2} ||d\theta + n ||^2 \qquad, \qquad S_1 
(\theta,\varphi, \ell)=
{\kappa \over 2}|| d\varphi + q \theta +  \ell||^2, \eqno(2.6)
$$

where $q$ is the charge of the matter field; it is an integer.

Alternatively, we can use the ``Wilson formulation":

$$
S_0 (\theta) ={\beta\over 2} \sum_p(1- \cos \ d \theta)_p \qquad, \qquad 
S_1 (\theta, \varphi) = {\kappa\over 2} \sum_{<ij>} 
(1 - \cos (d\varphi + q \theta)_{<ij>}).
\eqno(2.7)
$$

\underbar{Class C: Non abelian gauge theories coupled to matter fields 
breaking the gauge}

\underbar{group to a Cartan subgroup containing $U(1)$} .

As an example we analyze the Georgi--Glashow model (model C).

Its basic fields are an $SU(2)$--valued gauge field
$g$ and a Higgs  field $\Phi$, of unit length, in the adjoint representation of 
$SU(2)$. The action is given by

$$
S(g, \Phi) = S_0 (g) + S_1 (g, \Phi)
$$
$$
S_0 (g) = \beta \sum_p \Bigl(1 - \chi (g_{\partial p}) \Bigr) \quad, \quad
S_1 (g, \Phi) = \beta_H \sum_{<ij>}  \ (\Phi, U_H (g_{<ij>}) 
\Phi) \eqno(2.8)
$$

where $\chi$ denotes the character of the 
fundamental representation, and $U_H (\cdot)$ denotes the adjoint 
representation.

\vskip 0.3truecm
{\sl 2.2 Phase diagrams}
\vskip 0.3truecm

For small $\kappa$ and $\beta^{-1}$, model A has a 
Coulomb phase with a massless photon [13,12,14]. Furthermore, it is known 
that, for 
small $\beta$ and suficiently large $\kappa$, it has a superconducting 
Higgs phase [15].

Model B, with $q=1$, has a confining / Higgs phase, for $\beta$ 
small or $\kappa$ large . Furthermore one expects a massless phase 
for $\beta$ and $\kappa$ small [16].

For $q \not= 1$, a phase transition line is expected to separate the 
confining and the Higgs phases, furthermore, for $q$ sufficiently large, 
a Coulomb phase is known to appear [13,12] for intermediate 
values of $\beta$, for arbirary large $\kappa$. 
If we add a topological $\Theta$-term,

$$
{i\Theta \over 4\pi^2} q \sum_{c_4} [(d\theta + n) \wedge (d\theta + n)] 
(c_4), \eqno(2.9)
$$

where the wedge product on the lattice can be defined as in [17], to the 
action of model B, in the limit $\kappa \nearrow \infty$,
we recover the ${\bf Z}_q$ model discussed by 
Cardy and Rabinovici, which
is expected [11] to exhibit  Higgs, Coulomb, confinement and
``oblique confinement" phases , related to an (approximate) symmetry under 
an $SL (2, {\bf Z})$ duality group.

Model C has been rigorously shown to have a Coulomb phase 
for $\beta_H = \infty$ and $\beta$ large. The Coulomb phase is expected [18] 
to extend into a region of large values of $\beta_H$ and $\beta$.
A confining phase 
is also known to exist [15], for small values of $\beta$.

Charged particles and monopoles are expected to exist in the Coulomb phase of 
these models as (infra--)particle excitations. This is the main issue 
discussed in this paper.

For models in class C, magnetic monopoles are expected to exist
as massive particles also in the continuum 
limit (provided it exists). These are the celebrated at't Hooft--Polyakov 
monopoles.In the models of class B, monopoles are
a lattice version of the (singular) Dirac monopoles.

\vskip 0.3truecm
{\bf 3. Charges or monopoles}
\vskip 0.3truecm

To motivate our constructions of charged and monopole fields for the models
introduced in the previous section,  
we outline a general strategy exploited in [19,20,21,2](see also [22]) 
to construct 
charged (and soliton) fields and superselection sectors 
directly from correlation functions defined
through euclidean functional integrals.
We discuss this construction on the lattice; but similar ideas have been 
applied to continuum models, although only formally. 

\vskip 0.3truecm
{\sl 3.1 Reconstruction of charged and soliton quantum fields from 
euclidean Green functions }
\vskip 0.3truecm

Let ${\cal O}(C)$ denote a euclidean 
``observable", i.e., a neutral, gauge--invariant, local function of the basic 
fields with support on a compact connected set of cells $C$, such as a 
Wilson loop, a "string field", etc. 

We consider expectation values of such euclidean observables 

$$
G_m (C_1,...,C_m)= \langle {\cal O} (C_1)... {\cal O} (C_m) \rangle.
$$

Let $r$ denote reflection in the time--zero plane, $f$ a 
complex--valued function and $\bar f$ its complex conjugate. We define
the Osterwalder-Schrader (O.S.) involution $\Theta_{OS}$ as an 
antilinear map satisfying

$$
\Theta_{OS} f (\Phi_i) =\overline{f (\Phi_{ri})},
\Theta_{OS} f (g_{<ij>}) = \overline{f (g_{<r i r j>})},
$$
$$
\Theta_{OS} \Psi_i = \bar\Psi_{r i} \gamma_0, 
\Theta_{OS} \bar\Psi_i = \gamma_0
\Psi_{ri} \eqno(3.1) 
$$

Let ${\cal F}_+$ denote the set of linear combinations of euclidean 
observables ${\cal O}(C)$ with $C$ supported in the positive--time lattice. 
If, for every $F\in {\cal F}_+$,

$$
\langle (\Theta_{OS} F) F \rangle \geq 0 
$$

then the correlation functions $\{G_m\}$ are said to be O.S. positive.

\underbar{Theorem 3.1} (O.S. reconstruction). If the correlation functions $\{G_m
(C_1, ..., C_m) \}$ satisfy

\smallskip
i) lattice--translation invariance, and

ii) O.S. positivity

then one can reconstruct from $\{G_m\}$

a) a separable Hilbert space, ${\cal H}_0$ of physical states,

b) a vector of unit norm, $\Omega \in {\cal H}_0$, the vacuum,

c) a self--adjoint transfer matrix $T$, with unit norm, and unitary 
space translation operators $U_\mu, \mu=1, ... d-1$, acting on ${\cal 
H}_0$ and leaving $\Omega$ invariant.We define $T(t)=T^t,t \in {\bf Z}_+,
U({\vec a})= \prod_{\mu=1}^{d-1} U_\mu^{a_\mu},{\vec a}=(a_1,...,a_{d-1})
\in {\bf Z}^{d-1}$.

If, moreover, the correlation functions satisfy

iii) cluster properties

then

d) $\Omega$ is the unique vector in ${\cal H}_0$ invariant under $T$
and $U_\mu$.

\vskip 0.3truecm

From the explicit proof of the theorem it follows that there is a set of 
vectors $\{|C_1, ..., C_m > \}$ in ${\cal H}_0$, with $C_j$ contained in the 
positive time lattice, $j=1,...,m$,such that the set of linear combinations, 
denoted $\hat{\cal F}_+$, is dense in ${\cal H}_0$. 
On these vectors, the scalar product is defined by

$$
\langle C_1, ..., C_m| C^\prime_1, ..., C^\prime_n \rangle=
\langle \Theta_{OS} [O (C_1) ... O (C_m)] O(C^\prime_1)... O(C^\prime_n) 
\rangle \eqno(3.2)
$$

Field operators $\hat {\cal O} (C_1, ..., C_\ell)$, with $C_j$ 
contained in the strip 
$\{x : x^0 \in (0, t), t \in {\bf Z}_+ \}$, can be defined on $T(t) \hat{\cal
F}_+$ by setting

$$
\hat{\cal O} (C_1, ..., C_\ell) T(t) |C^\prime_1, ..., C^\prime_m > =|C_1, 
..., 
C_\ell, (C^\prime_1)_t, ...,(C^\prime_m)_t > \eqno(3.3)
$$

where $(\cdot)_t$ denotes translation by $t$ in the time direction. The 
operators $\hat {\cal O} (C)$ are the quantum mechanical operators
reconstructed from the
euclidean observables ${\cal O}(C)$. The algebra, ${\cal A}$, generated by the 
operators $\hat {\cal O}(C_1,...,C_\ell) T(t)$ defined above is called
"lattice observable algebra". 

If the space of the physical states of the model 
contains charged or soliton 
states, such states are not contained in the Hilbert space constructed above.

The construction underlying Theorem 3.1 reproduces only the vacuum sector of
such models. Suppose, however, that the Hilbert 
space,${\cal H}$, of physical states of the model can be decomposed into 
orthogonal sectors 
${\cal H}_q$ invariant under $T, U_\mu$ and the lattice observable algebra 
${\cal A}$, i.e.,

$$
{\cal H}= \oplus_q {\cal H}_q.
$$

Here ${\cal H}_0$ is defined to be the subspace given by $\overline{{\cal A} 
\Omega}$. 
It is called \underbar{vacuum sector}. 
A sector ${\cal H}_q \perp {\cal H}_0$ is 
called a \underbar{``charged sector"} if there are no lattice translation--
invariant vectors in ${\cal H}_q$.

To construct the charged sectors, suppose that
we can find an enlarged set of correlation functions $\{G_{n,m}\}$ obtained 
by taking expectation values involving,
besides euclidean observables, ``charged" 
order-- and disorder--fields, denoted by ${\cal C} (\Gamma)$ and by
${\cal D} (\Gamma)$, respectively, with 
support in a connected (but, in general, non--compact) set of cells 
$\Gamma$. 
Correlation functions with non vanishing total ``charge" $q$ are
obtained from those of vanishing total charge,by removing the support of 
the charge of a ``compensating" field of charge $-q$  to infinity. 

Assume that the correlation functions $\{G_{n,m} (C_1, ...,C_n, 
\Gamma_1, ..., \Gamma_m) \}$ still satisfy the hypotheses of the O.S. 
reconstruction theorem. 

Denote by ${\cal H}, T, U_\mu, \Omega, \hat{\cal O} (C), \hat {\cal C} 
(\Gamma), 
\hat{\cal D} (\Gamma)$ the Hilbert space, the transfer matrix, the 
translation operators, the vacuum and the quantum fields obtained via 
O.S. reconstruction from the correlation 
functions $\{G_{n,m}\}$.
If all the correlation functions of non-vanishing total ``charge"
vanish, and 
clustering holds, then the Hilbert space ${\cal H}$ contains 
charged sectors, because the scalar product between two charged states 
of unequal charge, defined in analogy with (3.2), vanishes; 
(for more precise definitions and proofs see [  ]).

The quantum fields $\hat{\cal C} (\Gamma), \hat{\cal D} (\Gamma)$ 
map the vacuum sector to a charged sector.

\smallskip
\underbar{Remark 3.1}
Some of the quantum fields might couple 
the vacuum to one--particle states, or infra--particle states. This can be seen 
as follows: Provided that $T \geq 0$,
the mass operator can be defined by

$$
M = - \ln (T\lceil \{{\cal H}^{(0)} \ominus {\bf C} \Omega \}) \eqno(3.4)
$$

where ${\cal H}^{(0)}$ denotes the fibre of ${\cal H}$ of zero total momentum,
(i.e. a generalized vector 
$|\Psi >$ belongs to ${\cal H}^{(0)}$ iff $U_\mu |\Psi > = |\Psi >$).
Then we have the following result.

\underbar{Theorem 3.2} A field operator $\hat A$ acting on
${\cal H}$ couples the vacuum $\Omega$ to a stable massive one--particle 
state iff

$$
< \hat A \Omega, T(t) U({\bf a})\hat A \Omega > - <\hat A \Omega, 
\Omega > <\Omega, \hat A \Omega >
\sim {e^{-m (\hat A) t} \over t^{{d-1\over 2}}}, \quad {\rm as} \  t\nearrow 
\infty \eqno(3.5)
$$

with $m(\hat A) > 0$, for any ${\bf a} \in {\bf Z}^{d-1}$. 
This decay law is called Ornstein--Zernike decay.(For infra--particles 
the exponent in the denominator on the r.h.s. of (3.5) has a positive 
small correction.)

For many lattice models, the large $t$ behaviour involved in (3.5) can be 
analysed in terms of expansions methods [23,19,21].

\vskip 0.3truecm
{\sl 3.2 Charged fields in non--compact abelian models}
\vskip 0.3truecm

In this section we recapitulate the basic steps of
the construction of charged fields in the models of class A [2,3], following 
the scheme outlined above. We do this for the sake of completeness and in 
order to elucidate 
the difficulties arising in attempts to extend our construction to 
models in class B.

Let $E$ be a real--valued 1--form with support on an infinite, connected 
sublattice of the time--zero hyperplane,$\Lambda_0$, of the lattice ${\bf Z}^4$.

Furthermore we assume that

$$
\delta E = \delta_0 \eqno(3.6)
$$

where 

$$
(\delta_0)_i = \cases{1  & if  $i=0$  \cr
0 & otherwise \cr} 
$$

and

$$
E_{<ij>} \sim d (<ij>, 0)^{-2}, {\rm as} \ \ d (<ij>, 0) \nearrow \infty
\eqno(3.7)
$$

Here $0$ is the origin in ${\bf Z}^4$ and  $d(<ij>, 0)$ denotes the 
euclidean distance between $<ij>$ and $0$. As an example, one may consider

$$
E = d_{\Lambda_0} \Delta_{\Lambda_0}^{-1} \delta_0 \eqno(3.8)
$$
 
We denote by $E(x)$ the 1-form $E$ translated by $x$; $E(x)$ 
describes the electrostatic Coulomb field surrounding a source 
of charge 1 located at $x$. We define the charged fields of the Higgs model 
by

$$
\Phi (x, q, E) = (\Phi_x)^q e^{i q (A, E(x))}, \quad
\Phi (x, -q, E) = (\bar\Phi_x)^q e^{-iq (A, E(x))}, \quad q \in {\bf N}
\eqno(3.9)
$$ 

(See fig.1)
Typical observables, $\cal{O}$, are Wilson loops 

$$
{\cal O} (\alpha C) = \prod_{<ij>\in C} e^{i A_{<ij>} \alpha} \eqno(3.10)
$$

where $C$ is a loop, $\alpha \in {\bf R}/ \{0\}$, and ``string fields"

$$
{\cal O} (C_{xy}) = \bar\Phi_x \prod_{<ij>\in C_{xy}} e^{iA_{<ij>}} \Phi_y
\eqno(3.11)
$$

where $C_{xy}$ is a path form $x$ to $y$.

Correlation functions $G^E_{n,m}$ are defined by 

$$
G^E_{n,m} (x_1 q_1, ..., x_n, q_n, C_1, ... C_m)=
\langle \prod^n_{i=1} \Phi (x_i, q_i, E) \prod^m_{j=1} {\cal O}(C_j) 
\rangle \eqno(3.12)
$$

For later purposes it is convenient to define a generalization of 
(3.12): For two different 1-form $E, E^\prime$ satisying (3.6), we set

$$
G_{n+m, r} (x_1, q_1, E, ..., x_n, q_n E, x^\prime_1, q^\prime_1, E^\prime, 
...,
x^\prime_m, q^\prime_m, E^\prime, C_1, ... C_r) =
$$
$$
\equiv \langle \prod^n_{i=1} \Phi (x_i, q_i, E) \prod^m_{j=1} \Phi 
(x^\prime_i, q^\prime_i, E^\prime) \prod_{\ell=1}^r {\cal O} 
(C_\ell)\rangle\eqno(3.13)
$$

The correlation functions $G_{n,m}$ can be expressed as sums over 
configurations of ``electric 
currents", by first integrating over the matter fields and then integrating 
over $A$. For example 

$$
G_{n,0}^E (x_1, q_1, ..., x_n, q_n)=
{\sum_{\rho: \delta(\rho+ \tilde E)=0} z(\rho) e^{-{\beta \over 2} 
(\rho+\tilde E, \Delta^{-1} (\rho + \tilde E))} \over \sum_{\rho : 
\delta\rho=0} z(\rho) e^{-{\beta\over 2}(\rho, \Delta^{-1} \rho)}}
\eqno(3.14)
$$

where $\rho$ is an integer valued 1-current, $z(\rho)$ is a certain statistical 
weight determined by the action of the model, and 

$$
\tilde E = \sum^n_{i=1} q_i E(x_i)
$$

Since, in the denominator of (3.18), $\delta\rho=0$, the support of the 
currents $\rho$ is given by a set of loops. Hence the denominator can be interpreted
as the partition function of a gas of current loops interacting via the 
four--dimensional (lattice) Coulomb potential $\Delta^{-1}$. In the 
numerator of (3.18), open currents appear, with sources 
at $\{x_i\}$, besides current loops. At the 
sources, the electric currents spread out in fixed time planes,
as described by $E(x_i)$. The currents $\rho$ can be interpreted as the 
Euclidean worldlines of charged particles;
currents supported on loops correspond to worldlines of virtual 
particle--antiparticle pairs, open currents with connected support 
correspond to the worldlines of  
particles created at one end of the line and annihilated at the other one.

\underbar{Theorem 3.3} The
correlation functions $G_{n,m}^E$ defined in (3.12) are lattice 
translation invariant and O.S. positive. Furthermore, for $\beta$ large 
enough or for strictly positive $\beta$ and $\kappa$ small enough, 
clustering holds  and all correlation functions
with non--zero total charge vanish.

\smallskip
\underbar{Idea of proof} 
Invariance under lattice translations follows from the
existence of the thermodynamic limit of the measure corresponding to 
(2.3) derived by 
correlation inequalities [15,16]. O.S. positivity of those measures 
can be proved as
in [15], and O.S. positivity of charged correlations follows from the fact 
that $E(x)$ is localized in a fixed--time plane, so that e.g.

$$
G^E_{2,0} (rx, - q, y, q) =
\langle \Theta_{OS} \bigl((\Phi_x)^q e^{iq(A, E(x))}\bigr) (\Phi_y)^q e^{iq 
(A, E(y))} 
\rangle 
$$

for $x^0, y^0 > 0$.(There is a slight subtlety cocerning the choice of 
boundary conditions for correlation functions of charged fields in a 
bounded space--time volume.It can be dealt with in a way similar to that 
explained in [21]; see also section 4.2)

Cluster properties are a consequence of 
(generalizations of) the bounds stated in the next theorem,for details 
see [2,3].

\underbar{Theorem 3.4} In model A, for $\beta$ large enough or $\kappa$ 
sufficiently small, and $|x-y|$ large enough,

$$
G_{2,0} (x, -1, E, y, 1, E^\prime) \leq {\rm exp}\{-c_1(\beta,\kappa) 
(\tilde E,\tilde E)\}
$$
$$
\leq {\rm exp} \ \{- \Bigl({\beta \over 2} - c_2(\beta,\kappa) \Bigr)
\Bigl((\tilde E - E_{dip}), \Delta^{-1} (\tilde E - E_{dip}) \Bigr) \}
 c_3(\beta,\kappa)^{|x-y|}  
\eqno(3.15)
$$

and for $x=ry$

$$
{\rm exp} \ \{- {\beta \over 2} ((\tilde E + \rho_{min}), \Delta^{-1} 
(\tilde E + \rho_{min})) \} c(\kappa)^{|x-y|} \leq G_{2,0}
(x, -1, E, y, 1, E^\prime)
\eqno(3.16)
$$

where $ c_i(\beta,\kappa)$ tends to 0 exponentially, as $\beta \nearrow 
\infty$, and, for fixed $\beta$, $c_i(\beta,\kappa) \sim O(\kappa)$, as 
$\kappa \downarrow 0$, $i=1,2,3$, and $c(\kappa) \leq 1$, 
$c(\kappa)\sim O(\kappa)$, as $\kappa \downarrow 0$ 
; $\rho_{min}$ is a current of minimal 
length and flux 1 connecting $x$ to $y$, $\tilde E = E(x) - E^\prime (y)$ and

$$
E_{dip} = d \Delta^{-1} (\delta_x - \delta_y) \eqno(3.17)
$$

The upper bound in Theorem 3.2 shows that charged (infra--)particles in 
the Higgs model have strictly positive mass in the range of coupling 
constants indicated in the theorem. The lower bound proves that, in the 
same range of coupling constants, their mass is finite.

The method of proof is based on a  combination of a Peierls-- and 
renormalization group 
argument, following [12,24]. For $\kappa=\infty$ the lower bound in (3.16) 
can be obtained more easily from Jensen's inequality, 
using representation (3.14).

\smallskip
\underbar{Remark 3.2} It can be shown, following [25], that, for $\beta$ 
sufficiently small and $\kappa$ sufficiently 
large, clustering holds, and

$$
G_{2,0} (x,1, E, y, -1, E^\prime) \geq e^{-O(\beta)( \tilde E, (\Delta + 
O(\beta))^{-1})\tilde E)} > {\rm const}. \eqno(3.18)
$$

uniformly in $|x-y|$. Hence correlation functions of non zero total 
charge do not vanish. This is a manifestation of charged--particle 
condensation typical for the Higgs phase.

\vskip 0.3truecm
From the O.S. reconstruction theorem  we obtain a Hilbert space, denoted by
${\cal H}(E)$, a transfer matrix $T_E$ and a dense set of vectors:

$$
|x_1, q_1, ..., x_n, q_n, C_1, ..., C_m >_E, \quad q_i \in {\bf Z}\backslash 
\{0\} 
\eqno(3.19)
$$

corresponding to charged fields of charges $\{q_i\}$ inserted at points 
$\{x_i\}$ and local observables located at $\{C_j\}$, with $x_i$ and $C_j$ 
in the positive time lattice. 

From (generalizations of) the lower bounds (3.16) it follows that 
$||T_E|| \not=0$, i.e. the states (3.19) have finite energy.

One can define non--local charged fields by

$$
\hat\Phi (\vec x, q, E) |x_1, q_1, ... x_n, q_n, C_1, ..., C_m >_E=
|x, q, x_1, q_1, ..., x_n, q_n, C_1, ..., C_m >_E \eqno(3.20)
$$

with $x^0 < x^0_1 <...< x^0_n$.

We now consider the region of coupling constant space where all correlation 
functions of non--zero total charge vanish. Then ${\cal H} 
(E)$ decomposes into orthogonal sectors labelled by the total electric 
charge $q$:

$$
{\cal H} (E) = \oplus_q {\cal H}_q (E).
$$

Cluster properties show that the sectors ${\cal H}_q (E)$, $q\not =0$, are 
charged sectors, in the sense described in sect.3.1. 

From the above construction it is clear that, a priori, the charged sectors
${\cal H}_q (E)$ depend on the choice of the distribution $E$.
It is natural to ask  if ${\cal H}_q (E)$ is orthogonal to ${\cal H}_q 
(E^\prime)$, for $E \not = E^\prime$. In order to answer this question , one
considers the generalized correlation functions (3.16).

One can define a scalar product between 
states in ${\cal H}_q (E)$ and ${\cal H}_q (E^\prime)$ by

$$
\quad_E <x, q| x^\prime, q >_{E^\prime} = G_{2,0} (rx, -q, E, x^\prime, q, 
E^\prime)
$$

From generalizations of the upper bounds (3.19), (3.20 ) it follows that 
${\cal H}_q (E) \perp {\cal H}_q (E^\prime) \ {\rm if}$

$$
\Bigl(( E - E^\prime), \Delta^{-1} (E-E^\prime) \Bigr) \eqno(3.21)
$$

diverges and one easily realizes that this divergence occurs if $E$ and
$E^\prime$ do not have the same ``behaviour at infinity".

There is an interesting choice of $E$ and $E^\prime$ which naturally leads 
to a divergence in (3.21). For example, if one chooses $E$ to be supported in a
spatial cone 
${\cal S}$ with apex in 0 and opening solid angle less than $4\pi$, the 
states obtained via O.S. reconstruction are the lattice approximation 
of the states discussed by Buchholz [26] in the algebraic approach to Q.E.D.. If
the field $E^\prime$ is chosen to be localized in a disjoint spatial cone
${\cal S}^\prime$, then (3.21) diverges. In particular, if we choose ${\cal 
S}^\prime$ to be the cone obtained by a rotation of ${\cal S}$ then this 
divergence shows that in the continuum limit (if it exists) the rotations 
cannot be unitarily implemented on ``Buchholz states". For more details 
see [2,3,19].

We conclude with some remarks about particle structure analysis on charge
$\pm 1$ sectors. By ispection of the proof of Theorem 3.2 (see [2,3]),
one can argue that for $|x-y|$ large :

$$
G^E_{2,0} (x, -1, y, 1) \sim \sum_{\rho_{xy}:\delta\rho_{xy} 
= \delta_x - \delta_y} z(\rho_{xy}) \int d\mu_{ren}
(A) e^{i(A,\rho_{xy} +\tilde E)} \eqno(3.22)
$$

where $\rho_{xy}$ is a current of flux 1 and connected support 
$\tilde E = E(x) - E(y),
z (\rho_{xy})$ is a statistical weight and $d\mu_{ren} (A)$ is a positive 
measure of the form

$$
d\mu_{ren} (A) = {1\over Z_{ren}} d\mu_{\beta_{ren}} (A) e^{{\cal I} (dA)}
\eqno(3.23)
$$

In (3.23), $d\mu_{\beta_{ren}} (A)$ is a gaussian measure with mean 0 and 
covariance $\beta_{ren} (\delta d)^{-1}$ (+ gauge fixing), where 
$\beta_{ren}$ is a renormalized coupling constant $[\beta_{ren} (\kappa,\beta) 
= \beta + O(c(\kappa) )]$ and ${\cal I} (dA)$ is a sum of gauge--invariant 
``irrelevant" terms, in the jargon of the renormalization group. Hence the 
large distance behaviour of $G^E_{2,0}$, which is independent of the 
irrelevant terms, ${\cal I}$, should essentially be given by a product of 
two factors: one is due to the fluctualing current line $\rho_{xy}$, and, 
from the analysis in terms of excitation expansions of [23], it is expected 
to produce an Ornstein--Zernike decay

$$
|x-y|^{-3/2} e^{-m (\kappa,\beta) |x-y|}
$$

corresponding to a particle of mass $m (\kappa,\beta) \sim -\ln c(\kappa)+ 
O(\beta)$.
The second factor can be argued to contribute another power correction 
to the exponential law

$$
{\rm exp} [- {\beta_{ren} \over 2} (\tilde E, \Delta^{-1} \tilde E)] \sim 
|x-y|^{-c \beta_{ren}}, c > 0
$$

It is due to the soft photons accompanying an infra--particle. 
Therefore, as $x^0 \nearrow \infty$, one expects that

$$
< \phi (0, \pm 1, E) \Omega, \phi (x, \pm 1, E) \Omega >
$$
$$
\sim {e^{-mx^0} \over(x^0)^{3/2+ c\beta_{ren}}}, \quad x^0 \nearrow \infty
\eqno(3.24)
$$

Equation (3.24)  exhibits the infraparticle nature of the 
charged particles in the Higgs model. In fact, the 
vacuum expectation value of a charged field vanishes in the region of 
coupling constant space considered, 
and a comparison with the general formula (3.5) shows
that the mass operator $M$, as defined in (3.4), does not have a discrete 
eigenvalue corresponding to a sharp one--particle state.

\vskip 0.3truecm
\underbar{Remark 3.3 (Q.E.D.)}
In lattice Q.E.D., the electron field is defined by

$$
\Psi (x,1, E) = \Psi_x e^{i(A, E(x))} 
$$

and the conjugate euclidean field is

$$
\Psi (x, - 1, E) \equiv \bar\Psi_x \gamma_0 e^{-i (A, E(x))}
$$

Typical local observables are Wilson loops and string 
fields. Correlation functions of charged fields and local observables 
satisfy the hypotheses of the Reconstruction Theorem 3.1 and,
for $\kappa$ sufficiently small, bounds analogous to those in Theorem 
[5~3.3 (as discussed in [2]).

This permits one to reconstruct non--local electron--positron field operators

$\Psi_\alpha (\vec x, q, E), q = \pm 1$, which are defined by

$$
\hat\Psi_\alpha (\vec x, q, E) |x_1, q_1, \alpha_1, ..., x_n q_n \alpha_n, C_1, 
... C_m >_E =
$$
$$
= |x, q, \alpha, x_1, q_1 \alpha_1, ..., x_n, q_n \alpha_n, C_1, ..., C_m 
>_E 
$$

for $x= ({1\over 2}, \vec x), x \not = x_j$.

The field $\Psi_\alpha (\vec x, q, E)$ can be viewed as the lattice 
approximation of the formal operator (1.1) introduced by Dirac. An analogue 
of equation (3.24) is expected to hold for $\kappa$ small enough, on the 
basis of the proof of bounds on electron-positron 
correlations. It would exhibit the infraparticle nature of electrons and 
positrons.

\vskip 0.3truecm
{\sl 3.3  Monopoles and duality}
\vskip 0.3truecm

Model B, in Villain form, at $\kappa=0$ is dual to model A in the limit
$\kappa\nearrow \infty$ and it describes an 
interacting theory of ``photons" and Dirac monopoles. Since monopoles can 
be viewed as solitons in such a model, 
we may appeal to the strategy outlined in sect. 3.1 to construct a monopole 
field operator by introducing  disorder fields, whose expectation values are 
the Euclidean Green functions of the Dirac monopoles [2,3,19]. 
For this purpose we 
introduce a real 3--form $B$ on the lattice ${\bf Z}^4$, given by 
$B=2\pi ^*E$ and define the disorder field by

$$
D_\omega (x_1, q_1, ..., x_n, q_n, B)= {e^{-{\beta\over 2} \{||d\theta + n + 
\delta \Delta^{-1} (\tilde B - \omega)||^2 -|| 
d\theta + n ||^2\}}} \eqno(3.25)
$$

where $x_i \in {\bf Z}^4, q_i \in {\bf Z} \backslash \{0\},$

$$
\tilde B = \sum_i q_i B (x_i)
$$

and $\omega$ is a $2 \pi {\bf Z}$--valued 3--form satisfying

$$
d(\tilde B -\omega) =0. \eqno(3.26)
$$

We now explain why expectation values of $D_\omega$ are correlation 
functions of monopoles.

First, we use the Hodge decomposition to rewrite

$$
n= d\Delta^{-1} \delta n + \delta \Delta^{-1} m \eqno(3.27)
$$

where

$$
m= dn \eqno(3.28)
$$

Let $n[m]$ be an integer--valued solution of the cohomological equation 
(3.28).
Then every other solution is of the form  $n[m]+ d\ell$, where $\ell$ is a
to $2\pi {\bf Z}$-valued 1-form. We define a real-valued gauge field $A$ 
by setting

$$
A= \theta + \ell + \delta \Delta^{-1} n [m] \eqno(3.29)
$$

and one can easily verify that

$$
\langle D_\omega (x_1, q_1, ..., x_n, q_n, B) \rangle =
$$
$$
={\sum_{m:dm=0} \int \prod_{<ij>} dA_{<ij>} e^{-{\beta\over 2}
|| dA+\delta \Delta^{-1}(m+ \tilde B-\omega)||^2} \over 
\sum_{m: dm =0} \int \prod_{<ij>} dA_{<ij>} e^{-{\beta\over 2} 
||dA+\delta \Delta^{-1} m||^2}}
\eqno(3.30)
$$

$A$ is the euclidean ``photon field", and the Hodge--dual of 
$(m-\omega)$ is supported on a set of lines in the dual lattice 
${\bf Z}^4_{1\over 2}$, which 
can be interpreted as Euclidean worldlines of Dirac monopoles; $\omega$ 
itself can be viewed as a Dirac string if we take

$$
\omega = \sum_i \omega_{x_i} + \omega_\infty
$$

where $^*\omega_{x_i}$ has support in an open line at constant time with one 
end at $x_i$ and the other end joining a compensating current 
$\omega_\infty$ at infinity (monopole b.c.).

Then $B(x_i) - \omega_{x_i}$ is exactly the lattice approximations of the 
magnetic field of a monopole located at $x_i$, together with its Dirac 
string $\omega_{x_i}$.

In the presence of the disorder field $D_\omega$ the Euclidean observables 
of the model must be modified so that expectation values do not depend on 
the choice of the ``Dirac string" $\omega$. In particular the Wilson loop 
${\cal O}(C)$ is now replaced by

$$
{\cal O}_\omega (S) ={\cal O} (C) \prod_{p\in S:\partial S=C} 
e^{-i (\delta \Delta^{-1} \omega)}\eqno(3.31)
$$

where $S$ is a surface; see [19,2].

Correlation functions in the $U(1)$ gauge theory are then defined by

$$
G^B_{n,m} (x_1, q_1, ..., x_n, q_n; S_1, ..., S_m)=
\langle D_\omega (x_1, q_1, ..., x_n, q_n, B) \prod^m_{i=1} {\cal O}_\omega
(S_i) \rangle \eqno(3.32)
$$

By duality e.g.

$$
\langle D_\omega (x_1, q_1, ..., x_n, q_n, B) \rangle = 
{\lim\limits_{\kappa \nearrow\infty}} \langle \prod^n_{i=1} 
\Phi (x_i, q_i, E) \rangle_\kappa
\eqno(3.33)
$$

where $\langle \cdot \rangle_\kappa$ denotes the expectation value in the Higgs
model A and $E= ^*B$. O.S. reconstruction theorem applied to 
$G^B_{n,m}$ provide us with non--local Dirac monopole fields ${\cal M} 
(\vec x, q,
B)$ and, for $\beta$ large enough, Dirac monopole sectors ${\cal H}_q(B)$. 
Furthermore a particle--structure analysis along the line of sect. 3.2 
exhibits the infraparticle nature of the Dirac monopole of charge $\pm 1$.

\vskip 0.3truecm
\underbar{Remark 3.3}
The monopole construction outlined above can be applied to every 
$U(1)$--gauge theory without matter fields.
This constructions and variants thereof have been applied in numerical
simulations of lattice theories  in [27], with the aim 
of detecting phase transitions between phases corresponding to different 
behaviour at large distances of the monopole Green functions, which has 
been described in the dual picture in Theorem 3.3 and Remark 3.2.

\vskip 0.3truecm
{\bf 4. Charges and monopoles}
\vskip 0.3truecm
We start this section with an outline of the new problems appearing in an 
attempt to extend the
construction sketched in the previous sections to models where charges and 
monopoles coexist (class B,C).

\vskip 0.3truecm
{\sl 4.1  \ Dirac quantization condition}
\vskip 0.3truecm

We have seen that, for $\kappa=0$, model B (in Villain form) can be  
written explicitly in terms of a ``photon field" $A$ and a ``monopole field"
$^*m$. The Dirac strings of the virtual monopole loops in the partition
function sweep out surfaces described by the support of $^*n$. 
The term $S_1$ in the
action introduced in eq.(2.6) couples the gauge field $\theta$ to a 
charged matter field. In our expression for the partition function, the 
charged matter gives rise to charged loops describing the worldlines of 
virtual particle--antiparticle pairs. We must ask whether the 
statistical weights of these charged loops depend on the location of the 
Dirac strings of virtual monopoles. 

To answer this question, let us rewrite the action (2.6) in terms of $A,m, 
\varphi, \ell$. We obtain, for $q=1$,

$$
S=  {\beta \over 2} || dA||^2 + {\beta \over 2} (m, \Delta^{-1} m) 
+ {\kappa\over 2} || d\varphi + A - \delta \Delta^{-1} n [m] + \ell ||^2 
\eqno(4.1)
$$

Independence of the Dirac strings corresponds, as recalled above, to 
independence of the choice of the two--form $n[m]$ satisfying 
$d n [m]= m$, as in (3.28).

Let $\zeta$ be a ${\bf Z}$-valued 1-form. Then by Poisson summation
formula,

$$
\sum_\ell e^{-{\kappa \over 2} || d\varphi +A -\delta \Delta^{-1} n [m] + 
\ell ||^2} =
\sum_\zeta e^{-{1\over 2\kappa} ||\zeta||^2} e^{i(\zeta, d\varphi + A - \delta 
\Delta^{-1} n [m])}
$$

Integrating over $\varphi$ we obtain

$$
\delta \zeta =0.
$$

We conclude, using the Poincar\'e lemma, that there exists a ${\bf 
Z}$-valued 2-form, $\chi$, such that

$$
\zeta = \delta \chi,
$$

so that

$$
(\zeta, \delta \Delta^{-1} n [m]) = (\delta \chi, \delta \Delta^{-1} n [m])
=(\chi, d \delta \Delta^{-1} n [m]) = (\chi, n [m]) - (\chi, \delta 
\Delta^{-1} m)
$$

where the last equality follows from Hodge decomposition.

The second term depends only on $m$ and 

$$
e^{i (\chi, n[m])} =1.
$$

This is nothing but the Dirac quantization condition, since the electric 
charges appearing in the partition functions are $q_e \in {\bf Z}$ and the 
magnetic charges are $q_m \in 2\pi {\bf Z}$, so that

$$
q_e q_m \in 2\pi {\bf Z} \eqno(4.2)
$$

From the above proof it is clear that if we consider the expectation value 
of $e^{i(\theta, E)}$, where the ``electric current" 1-form $E$ is 
\underbar{not} integer valued, as introduced in 
sect 3.2, we would encounter an inconsistent dependence on the choice of 
Dirac strings. This problem is often neglected in the physics 
literature, wher virtual monopole loops are sometimes ignored, assuming that 
monopoles are ``very heavy", see e.g. [28]. But, in order to arrive at 
a fully consistent definition of 
gauge--invariant charged field Green functions, one must cope with it.

The natural suggestion is to replace the electric field $E (x)$ of sect. 
3.2 by an integer--valued electric current, a ``Mandelstam string", starting 
at $x$ and ending at the location of some compensating 
charge which will eventually be sent to infinity.

A naive idea would be to simply replace the variable $\phi_x e^{i(A, E(x))}$, 
used in the construction of charged states in model A, by 

$$
c_x^R e^{i\varphi_x} e^{i(\theta, \gamma^R_x)} \eqno(4.3)
$$

in model B, and then take the limit $R\nearrow \infty$. In (4.3), 
$\gamma_x^R$ is a unit 1-form with support on a straight line in a 
fixed--time 
plane from $x$ to some point at a distance $R$ and then joning that 
point to a fixed point in the 
time--zero plane, and 
$c_x^R$ is some normalization factor.

Consider the 2-point function: Integrating out $\varphi$ we obtain

$$
\langle e^{i\varphi_x} e^{i(\theta, \gamma_x^R)} e^{-i \varphi_y} 
e^{-i(\theta, \gamma_y^R)} 
\rangle =
{1\over Z} \int \prod_{<ij>} d\theta_{<ij>} e^{-S_0(\theta)} 
\sum_{\rho:\delta\rho=\delta_x-\delta_y} z (\rho) e^{i(\theta, \gamma_x^R
- \gamma^R_y + \rho)} \eqno(4.4)
$$

where $\rho$ are ${\bf Z}$-valued currents whose statistical 
wheight is denoted by $z(\rho)$, and $Z$ is the partition function. The phase 
factors appearing in (4.4) define Wilson loops, 
and, in the region of coupling constant space
where monopoles are particle excitations, 
the expectation value of the Wilson loop is known to exhibit perimeter
decay. Hence we expect (4.4) to vanish exponentially fast as $R\nearrow 
\infty$.

This decay is dominantly due to the self--energy of the strings $\gamma$.
This effect could eliminated by adjusting $c$. 
But we would then be left with an
interaction term between the strings which, being attractive and extending over 
the full string, tends to infinity in the limit $R\nearrow \infty$, and 
this appears to render the renormalized Green function divergent. Furthermore, 
since the interaction term depends on the time distance between strings, it 
cannot be renormalized away without violating O.S. positivity.
This problem could be solved if we replace a ``straight Mandelstam 
string" by a sum over ``fluctuating Mandelstam strings",
weighted by a measure which is concentrated on strings 
fluctuating so strongly that, with probability one, their ``interaction 
energy" remains finite
in the limit $R\nearrow \infty$.

A proposal for a ``natural" measure can be inferred from a representation
of correlation functions of a $U(1)$ scalar field, $\chi$, 
with coupling constant 
$\beta_\chi$, coupled to an external $U(1)$ gauge field $\theta$, 
in terms of random 
walks. E.g., in  $d$ dimensions, with $< \cdot > (\theta)$ denoting the 
corresponding expectation value,

$$
\langle e^{i\chi_x} e^{-i \chi_y} \rangle (\theta) =
\sum_{\omega_{xy}} {(2d)^{-|\omega_{xy}|-1} \over \beta} {Z (\theta |
\omega_{xy}) \over Z(\theta)} e^{i(\omega_{xy}, \theta)} \eqno(4.5)
$$

where $\omega_{xy}$ is a path (``string") from $x$ to $y$, 
$|\omega_{xy}|$ its length, $Z(\theta)$ is the partition function of the 
system, and $Z(\theta|\omega_{xy})$ is a partition function modified by 
$\omega_{xy}$, see [29].

In $d\geq 3$ dimensions, for $\beta_\chi$ sufficiently large and 
$\theta=0$, the string 
$\omega_{xy}$ is known to be rough. In fact 

$$
\langle  e^{i\chi_x} e^{-i\chi_y} \rangle (0) \longrightarrow {\rm const} 
$$

as $|x-y| \rightarrow \infty$, and it is believed that it remains in the 
rough phase for a class of positive measure of $U(1)$-external gauge fields, 
(roughly speaking, if $d\theta$ is sufficiently small in average).

This suggests that two--point functions of a $U(1)$ scalar field might yield 
an appropriate measure on ``Mandelstam 
strings", if $\beta_\chi$ is large enough.

For a better understanding of our construction in the compact models, we 
show how to reproduce the results already obtained in the non--compact models 
following the above ideas.

\vskip 0.3truecm
{\sl 4.2 \ The non--compact model revisited}
\vskip 0.3truecm

Let $\Gamma(R,t),R \in {\bf Z}_+,t \in {\bf Z}_{1/2}$, be a cube of height 
$2 |t|$ in the time ($x^0$--) direction and with sides of length $R$ in 
coordinate planes $x^0=$const and centered at the origin; see fig.2. We define 
$\Lambda(R,t)$ to be given by $\partial \Gamma (R,t) \cap \{x:x^0 \mma 0 
\}$, for $t \mma 0$.(To simplify our notation, the restriction of a 
lattice operator to $\Lambda(R,t)$ will be denoted with a 
subscript $\Lambda$, the specific sublattice which we are referring to 
being identified from context.)
We introduce a real-valued $0$-form $\lambda$ on $\Lambda(R,t)$, with action
given by

$$
S_\Lambda (\lambda) = {\beta_\lambda\over 2} || 
d_\Lambda \lambda + A ||^2_\Lambda \eqno(4.6)
$$

where $(,)_\Lambda$ is the inner product in $\Lambda(R,t)$, and 
$||f||_\Lambda^2=(f,f)_\Lambda$.

Denote by $\langle \cdot \rangle_{\Lambda(R,x^0)} (A)$ the (normalized) 
$\lambda$--expectation corresponding to the action $S_\Lambda(\lambda)$.
Set $R_\pm = (\pm {1\over 2}, R, 0, 0)$ and, 
for $x^0 > 0, y^0 < 0$, define

$$
G(x, y) = {\lim\limits_{R\rightarrow\infty}} \langle \langle e^{i\lambda_x}
e^{-i\lambda_{R_+}} \rangle_{\Lambda(R,x^0)} (A)
\langle e^{-i\lambda_y} e^{i\lambda_{R_-}} \rangle_{\Lambda(R,y^0)} (A)
e^{iA_{< R_+R_->}} \bar\Phi_x \Phi_y \rangle \eqno(4.7)
$$

where $\langle \cdot  \rangle$ is the expectation value 
of model A.

By explicit computation

$$
\langle e^{i\lambda_x} e^{-i\lambda_ {R_+}} \rangle_{\Lambda(R,x^0)} (A)=
e^{-{1\over 2 \beta_\lambda} ((\delta_x - \delta_{R_+}), 
\Delta_\Lambda^{-1} (\delta_x - \delta_{R_+}))} e^{i(E_\Lambda (x)
- E_\Lambda (R_+), A)} \eqno(4.8)
$$

where

$$
E_\Lambda (x) = d_\Lambda \Delta_\Lambda^{-1} \delta_x \eqno(4.9)
$$

Hence

$$
G (x,y) ={\lim\limits_{R\rightarrow \infty}} \langle \Bigl[e^{i( E_\Lambda (x)
- E_\Lambda (y), A)} \bar\Phi_x \Phi_y \Bigr]
[e^{i(E_\Lambda (R_-) - E_\Lambda (R_+) +\delta_{ < R_+ R_->}, A)} ]\rangle
$$
$$
=c \langle e^{i(E(x) - E(y), A)} \bar\Phi_x \Phi_y \rangle
\eqno(4.10)
$$

where $\delta_{<R_+ R_->}$ denotes the 1--form with support on the link
$<R_+ R_->$, whose value is 1 on that link, and

$$
c= {\lim\limits_{R\rightarrow \infty}} \langle 
[e^{i(E_\Lambda (R_-) - E_\Lambda (R_+) +\delta_{ < R_+ R_->}, A)} ]\rangle.
$$

To prove (4.10), we note that the form $E_\Lambda(x)-E_\Lambda(y)$ decays 
like $d^{-3}$ in the limit $R\nearrow \infty$, and $E_\Lambda(x) \rightarrow
E(x)$, where $E$ is given in (3.8).

We observe that the two--point function of the auxiliary matter 
field $e^{i\lambda}$ exactly reproduces the exponential $e^{i(A,E)}$ needed 
in the construction of charged states. Furthermore our construction 
respects O.S. positivity and, when the limit $R\nearrow \infty$ is taken, 
lattice translation invariance is restored.

Sectors corresponding to different choices of the one--form 
$E$ can be reproduced by modifying $ S_\Lambda (\lambda)$. 

\vskip 0.3truecm
{\sl 4.3  Charged fields in compact models}
\vskip 0.3truecm

It is natural to generalize the construction of the previous 
section to the compact model B.

We introduce a $U(1)$-valued scalar field $\chi$ and a $2\pi {\bf Z}$-valued 
1--form $r$ on the lattices $\Lambda(R,t)$ defined in the previous 
section. An action functional for $\chi$ and $r$ is defined by

$$
S_\Lambda (\chi, r) = {\beta_\chi \over 2} || d_\Lambda \chi - \theta + 
r ||^2_\Lambda \eqno(4.12)
$$

We denote by $\langle \cdot \rangle_{\Lambda(R,t)} (\theta)$ the expectation value 
w.r.t. the action (4.12).

With notations analogous to those of the previous section, we propose 
the following definition of the two--point function of the charge--1 field
in model B:

$$
\eqalign
{
&G(x,y) =\cr
&{\lim\limits_{R\rightarrow\infty}} c^R_{x,1} c^R_{y,1} 
\langle\langle e^{-i\chi_x}
e^{i\chi_{R_+}} \rangle_{\Lambda(R,x^0)} (\theta)
\langle e^{i\chi_y} e^{-i\chi_{R_-}} \rangle_{\Lambda(R,y^0)} (\theta) 
e^{i\theta_{<R_+R_->}} e^{i\varphi_x - i\varphi_y} \rangle\cr
}
\eqno(4.13)
$$

(See fig.2) 
In (4.13) $c_{x,q}^R$ is a normalization factor given by

$$
c_{x,q}^R = \{\langle\langle e^{-i\chi_{x_+}} e^{i\chi_{R_+}} 
\rangle_{\Lambda(R,1/2)}(\theta) \langle e^{i\chi_{x_-}}
e^{-i\chi_{R_-}} \rangle_{\Lambda(R,-1/2)}
(\theta)e^{-i\theta_{<R_+ R_->}} e^{i\theta_{<x_+ x_->}} \rangle
\}^{-{1\over 2}} \eqno(4.14)
$$

where $q$ is an integer and $x_{\pm} = (\pm {1\over 2}, \vec x).$ An 
approximate evaluation of (4.14) suggests 
that $c_{x,q}^R$ might be bounded in $R$ (see Appendix A).

The charge--1 two--point function $\langle e^{i\chi_x} 
e^{-i\chi_{R_+}}\rangle_{\Lambda(R,x^0)} (\theta)$ is periodic in 
$\theta$, with period 1. Hence, it has a Fourier representation 

$$
\langle e^{i\chi_x} 
e^{-i\chi_{R_+}}\rangle_{\Lambda(R,x^0)} (\theta)=
\int d\mu_{R_+} (j_x) e^{i (j_x,\theta)}
\eqno(4.15)
$$

where $j_x$ are ${\bf Z}$--valued one--forms satisfying

$$
\delta j_x=\delta_x-\delta_{R_+}
$$

and $d\mu_{R_+}(j_x)$ is a complex measure on the space of 1--forms $j_x$.
One can integrate out the matter field $\varphi$ in (4.13) and express 
the contribution as a weighted sum of ${\bf Z}$--valued 1--currents. In 
the numerator of (4.13), a contribution due to a 1--current $\omega$ 
satisfying

$$
\delta \omega=\delta_x-\delta_y
$$

will appear. As a result of this representation of expectation values in 
terms of sums over integer--valued currents, equation (4.13) yields a 
representation of the Green function $G(x,y)$ as an expectation value of 
a weighted sum of Wilson loops

$$
W_\omega(j_x,j_y)=e^{i(j_x+j_y+\delta_{<R_+ R_->}+\omega,\theta)}
\eqno(4.16)
$$

(See fig.2)

To analyze $\langle e^{-i\chi_x} e^{i\chi_{{R_+}}} 
\rangle_{\Lambda(R,x^0)} (\theta)$,
we perform a change of variables analogous to (3.36). Define

$$
v= d_\Lambda r \eqno(4.17)
$$

and let $r(v)$ be a fixed $2\pi {\bf Z}$- valued solution of equation (4.17).
We then define a real--valued scalar field $\lambda$ by

$$
\lambda = \chi + \delta_\Lambda \Delta_\Lambda^{-1} r [v] 
$$

Changing variables from $\chi, r$ to $\lambda, v$  we obtain

$$
\langle e^{-i\chi_x} e^{i\chi_{R_+}} \rangle_{\Lambda(R,x^0)} (\theta) =
$$
$$
= {\sum_{v:dv=0} \int \prod_i d \lambda_i e^{-{\beta_\chi\over 2} || 
d_\Lambda \lambda + 
\delta_\Lambda \Delta_\Lambda^{-1} v + \theta ||^2_\Lambda} e^{i(\lambda -
\delta_\Lambda \Delta_\Lambda^{-1} r [v], \delta_x - \delta_{R_+})_\Lambda} 
\over \sum_{v:dv=0} \int \prod_i d\lambda_i
e^{-{\beta_\chi \over 2} || d_\Lambda \lambda + \delta_\Lambda 
\Delta_\Lambda^{-1} v+ \theta ||^2_\Lambda}}
$$
$$
= e^{- {1\over 2\beta_\chi} ((\delta_x - \delta_{R_+}), \Delta_\Lambda^{-1} 
(\delta_x- \delta_{R_+}))} e^{-i(E_\Lambda(x,R_+), 
\theta)} F (E_\Lambda(x, R_+) |d\theta) \eqno(4.18)
$$

where

$$
F (E |d\theta) = {\sum_{v:dv=0} e^{-{\beta_\chi\over 2} (v+ d\theta, 
\Delta_\Lambda^{-1} (v+d\theta))_\Lambda} e^{-i(E_\Lambda, r[v])} 
\over \sum_{v:dv=0}
e^{-{\beta_\chi \over 2} (v+d\theta, \Delta_\Lambda^{-1} 
(v+d\theta))_\Lambda}}
\eqno(4.19)
$$

and

$$
E_\Lambda (x, R_+) = E_\Lambda (x) - E_\Lambda (R_+) .
$$

The action (4.12) describes a $U(1)$ spin system in the presence of the 
external gauge field $\theta$; the first two factors on the r.h.s. of (4.18) 
correspond to
the spin--wave (gaussian) approximation of the correlation function, the 
last one to the contribution of the vortices, described by $v$, to the spin 
system.

For $\beta_\chi$ sufficiently large and $d\theta=0$, the spin system is known 
to be in a phase with long range order, where the spin--wave approximation is
known to capture the essential physics at large distances. It is argued in [29],
as remarked above, that, in the presence of the gauge field, the system
remains in a phase with LRO, provided $d\theta$ is sufficiently ``small" in 
average. Hence, for $\beta_\chi,\beta$ sufficiently large, we expect the
vortices to form a dilute gas, so that the factor 
$F$ in (4.18) is expected to represent a ``small" correction to the spin--wave 
approximation. 
However, the presence of $F$ is crucial for the periodicity in $\theta$
which ensures independence of the Dirac strings of virtual monopole 
loops. 

Following the arguments of sect. 3.2 (but see also [2,14]), one can 
argue that, for $\kappa$ small enough, the contribution of the matter field at 
large distances essentially yields a renormalized $U(1)$ gauge theory with a
coupling constant

$$
\beta^\prime_{{\rm ren}} = \beta + O(\kappa^4)
$$

which, according to [12,24,3], and sect. 3.4, is expected to  renormalize 
to a non--compact gauge theory
with coupling constant

$$
\beta_{{\rm ren}} = \beta^\prime_{{\rm ren}} + e^{-O(\beta^\prime_{{\rm 
ren}})}
$$
 
in the scaling limit, provided $\beta^\prime_{ren}$ is large enough.
Accordingly, the 2--point function is expected to behave like

$$
G(x,y)\sim {1\over Z} \int {\cal D} A e^{-\beta_{{\rm ren}} 
||dA||^2} \sum_{\rho_{xy}:\delta\rho_{xy}=\delta_x - \delta_y}
z(\rho_{xy})
e^{i(A, \rho_{xy} + E(x) - E(y))}
$$
$$
F^G (E(x) | dA) F^G (-E(y) | dA), 
\eqno(4.20)
$$

at large distances , where $F^G$ is a 
gaussian approximation of $F$ estimated in Appendix A by

$$
F^G (E|dA) \sim e^{-[e^{-O (\beta_\chi)} ||E||^2 + e^{-O(\beta_\chi)} || 
E\cdot  \delta_\Lambda \Delta_\Lambda^{-1} dA||^2]} \eqno(4.21)
$$

Appealing to universality in the scaling limit, we believe that the large 
distance properties of $F$ are indeed correctly captured by the gaussian 
approximation, $F^G$.

Green functions with vanishing total charge
$G_{m+n}(x_1,q_1,...,x_m,q_m,y_1,q^\prime_1,...,y_n,q^\prime_n)$,
$x_i^0>0,y_i^0<0$, are defined by

$$
G_{m+n}(x_1,q_1,...,x_m,q_m,y_1,q^\prime_1,...,y_n,q^\prime_n)=
$$
$$
{\lim\limits_{R\rightarrow\infty}} 
\langle\prod_{j=1}^{m} c_{x_j,q_j}^R
 e^{-i q_j \chi_{x_j}}
e^{i q_j \chi_{R_+}}
\rangle_{\Lambda(R,x^0_j)} (\theta) e^{i q_j \varphi_{x_j}}
$$
$$
\prod_{k=1}^{n} c_{y_k,q^\prime_k}^R
\langle e^{-i q^\prime_k \chi_{y_k}} e^{i q^\prime_k \chi_{R_-}} 
\rangle_{\Lambda(R,y^0_k)} (\theta)  e^{i q^\prime_k \varphi_{y_k}} 
e^{i q \theta_{<R_+R_->}} \rangle,
\eqno(4.22)
$$

where $q=\sum_{j=1}^n q_j$ and $c_{c,q}^R$ is as in (4.14).

The large distance behaviour described by (4.20) implies cluster properties 
enabling us to 
reconstruct charged sectors and non--local charged field operators 
$\hat\Phi(\vec x,q,E)$ for model B. Equation (4.20) also suggests that
the particles of 
charge $\pm 1$ are infraparticles (see sect.3.2), since the contribution 
due to $F^G$ does not change qualitatively the large distance behaviour 
[30].
The same ideas also apply straightforwardly to  model B with $q\not= 
1$.

These observations suggest that the Green 
functions of charged fields in the present model have large--scale
properties similar to those in gauge theories 
\underbar{without} dynamical monopoles, constructed according to Dirac's 
original proposal, provided $\beta_\chi$ is chosen 
large enough. On short distance scales, the Dirac quantization condition 
implies integral quantization of electric flux accompanying a charged 
field $\hat\Phi(x,q,E)$ in all models with dynamical monopoles, while 
electric flux is \underbar{not} quantized in Dirac's construction. But, on 
large distance scales, the constraint of integer electric flux becomes 
"irrelevant", (i.e. this constraint renormalize to zero). The phenomenon 
described here is analogous to the phenomenon of symmetry enhancement 
studied in [12,31].

\smallskip

\underbar{Remark4.1}
The construction described above can be generalized 
to Green functions of charged fields in  deconfined (non--abelian 
Coulomb) phases of non--abelian gauge theories with some gauge group $G$,
by simply replacing the auxiliary two--point correlation functions of the
$U(1)$ scalar field 
$\chi$ coupled to the $U(1)$ gauge field by two--point correlation
functions of a non linear sigma model with target $G$ coupled to the gauge 
field. These correlation functions can be expressed [29,32] in terms of sums 
over path--ordered exponentials of the gauge field. It is 
useful to recall that an inequality analogous to 
the one quoted after (4.5) holds. If $G$ contains
a $U(1)$ subgroup and the field, $g$, of the non--linear sigma model, 
transforms under a representation, $U$, corresponding to a character 
non trivial on
this subgroup, then, at weak coupling, the 
two--point function in an external gauge field of zero curvature,
$\langle U(g_x) U(g^{-1}_y) \rangle (0)$ 
is bounded uniformly from below in $|x-y|$. One expects 
that this long range order persists in the model coupled to the 
$G$-gauge field, provided the curvature of the gauge field is small 
enough in average [29].

\vskip 0.3truecm
{\sl 4.4 Monopoles in compact models with matter}
\vskip 0.3truecm

In this section, we propose a construction of monopole Green functions 
in compact
models with matter (class B) obtained through a duality transformation.

First we identify the model dual to B. It can be described in terms of a 
${\bf Z}$-valued 2-form $\gamma$ and a 
${\bf Z}$-valued 1-form $\alpha$ (defined on the dual lattice) with an 
action given by

$$
S(\alpha, \gamma) = {1\over 2\beta} ||d\alpha - q\gamma||^2 + {1\over 
2\kappa} ||d\gamma||^2 \eqno(4.23)
$$

We can replace $\alpha$ by a real--valued 1--form $A$ by inserting the 
constraint

$$
\sum_m \delta (A - m) = \sum_\rho e^{i(A,\rho)}
$$

where $m$ is a ${\bf Z}$-- and $\rho$ is a $2\pi {\bf Z}$--valued 1--form,
and we have applied the Poisson summation formula.

The partition function of the dual model can then be written as

$$
Z = \sum_{[\gamma]} \int \prod_{<ij>} d A_{<ij>} 
e^{-{1\over 2\beta} || dA - q \gamma 
||^2 - {1\over 2\kappa} ||d\gamma ||^2} \sum_{\rho:\delta\rho=0} e^{i
(A,\rho)}, \eqno(4.24)
$$

where we exploit the gauge--invariance of the $A$ measure to impose the 
constraint $\delta\rho=0$.

The interpretation of formula (4.24) is clear: it describes an abelian 
gauge theory with charged matter fields and dynamical monopoles 
satisfying Dirac's quantization condition. The 1-form $\rho$ describes 
the worldlines of electric point charges (with values in $2 \pi 
{\bf Z})$, the 1-form $^*d \gamma $ describes the worldlines of magnetic 
point charges (which are integer--valued). This interpretation is 
consistent with the feature of duality transformations that they 
exchange electric and magnetic charge.

Accordingly, a Green function of charged fields in the dual model corresponds
to a monopole Green function in the original model. If we try to follow 
the construction of the previous section we encounter the following 
problem: we can reabsorb the ``Mandelstam strings" appearing in an 
expansion of the $\chi$--correlation functions, as in (4.5) and (4.13), 
in a shift of the current $\rho$. One may say that these ``Mandelstam 
strings" are ``screened" by the charge loops appearing in the dual 
model. Starting from the action (4.23), we meet the same phenomenon in 
the following way: Since the gauge field $\alpha$ is integer--valued and 
the electric charge of the auxiliary field $\chi$ is $2\pi$, the 
dynamics of $\chi$ is independent of $\alpha$, (i.e.,coupling $\chi$ to 
$\alpha$ does not have any physical effects).

A way out of this difficulty appears when $q\not=1$, because the
magnetic charges appearing
in the model are multiples of $q$, so that one can introduce electric 
loops of charge ${2\pi\over q}$, still satisfying Dirac's 
quantization condition.

Hence, instead of inserting correlation functions described as sums over 
a single fluctuating string of charge $2\pi$, one can introduce correlation 
functions described as sums over $q$ distinct fluctuating strings of charge
${2\pi\over q}$ which are \underbar{not} screened.

For this purpose we introduce a scalar field $\chi$ with value in $[-\pi q, 
\pi q]$ and a $2\pi {\bf Z}$--valued 1--form $r$ with action

$$
S_\Lambda(\chi, r) = {\beta_\chi\over 2} || { d_\Lambda \chi -2\pi A \over q} + 
r||^2_\Lambda \eqno(4.25)
$$

and denote the expectation value w.r.t. the action (4.25)
by $\langle \cdot \rangle_{\Lambda(R,t)} (A)$,
(see sect. 4.2). The charge--1 two--point 
correlation function $\langle e^{i\chi_x} e^{-i\chi_{R_+}} 
\rangle_{\Lambda(R,x^0)} (A)$
has a Fourier representation

$$
\langle e^{i\chi_x} e^{-i\chi_{R_+}} \rangle_{\Lambda(R,x^0)} (A) =
\int d \mu_{R_+} (j_x) e^{i(j_x,A)} 
\eqno(4.26)
$$

where $j_x$ are ${2\pi{\bf Z} \over q}$-valued 1--forms satisfying

$$
\delta j_x =2 \pi \delta_x - 2 \pi \delta_{R_+}
$$  

and $d \mu_{R_+}(j_x)$ is a complex measure on $j_x$. As a 2--point 
function of charged 
fields in the model dual to B, for $q \not=1$, we propose to consider

$$
G^* (x,y) = {\lim\limits_{R\rightarrow \infty}} c^R_x c^R_y \langle \langle
e^{-i\chi_x} e^{-i\chi_{R_+}} \rangle_{\Lambda(R,x^0)} (\alpha)
\langle e^{i\chi_y} e^{-i\chi_{R_-}} \rangle_{\Lambda(R,y^0)}
(\alpha) \rangle^*=
$$
$$
{\lim\limits_{R\rightarrow \infty}}c^R_x c^R_y \langle \langle e^{i\chi_x}
e^{-i\chi_{R_+}} \rangle_{\Lambda(R,x^0)} (A) \langle e^{i\chi_y}
e^{-i\chi_{R_-}}\rangle_{\Lambda(R,y^0)} (A) e^{i(\omega,A)}
e^{i 2\pi A_{<R_+R_->}} \rangle^*
\eqno(4.27)
$$

where $\langle \cdot \rangle^*$ is the expectation value of the dual 
model, and $\omega$ is a $2\pi {\bf Z}$--valued 1--form satisfying

$$
\delta\omega = 2\pi (\delta_x - \delta_y)\eqno(4.28)
$$

(See fig.3)
Note that $G^*$ is independent of the choice of $\omega$
satisfying (4.28).

The corresponding two--point function in the original compact model is 
obtained by applying an inverse duality transformation, by replacing the 
Wilson loops

$$
W_\omega (j_x,j_y)= e^{i(j_x + j_y +2\pi \delta_{<R_+ R_->} + \omega, A)}
$$

appearing in an expansion of (4.27) (as discussed in the previous 
section, see (4.16)) by 't Hooft
disorder loops.

To construct the 't Hooft loop, we have to identify a surface whose boundary 
is given by the loop $L= j_x + j_y +2\pi \delta_{<R_+ R_->} + ^*\omega$. 
Let us denote
its ${2\pi{\bf Z} \over q}$--valued Poincar\'e  
dual by $\Sigma (j_x, j_y; \omega)$. The 't Hooft disorder field, 
$D_\omega (j_x,j_y)$, is obtained by shifting 
$d\theta$ by $\Sigma$ in the original model, i.e.,

$$
D_\omega (j_x,j_y)= e^{-{\beta\over 2} (||d\theta +\Sigma 
(j_x, j_y; \omega) + n||^2 - || d\theta + n||^2)}
$$

Combining (4.26) and (4.27), the two--point monopole function in model B 
is given by

$$
G^*(x,y) = 
{\lim\limits_{R\rightarrow \infty}} c_{x,1}^R c_{y,1}^R 
\langle \int d \mu_{R_+} (j_x) d \mu_{R_-} (j_y)
D_\omega (j_x,j_y) \rangle \eqno(4.29)
$$

Equation (4.29) expresses $G^*(x,y)$ as an expectation value of a weighted 
sum of 't Hooft disorder fields $D_\omega(j_x,j_y)$.
In order to better understand the meaning of (4.29), notice that, 
as in sect.4.2, 
one can easily show that 

$$
\langle e^{i\chi_x} e^{-i\chi_{R_+}} \rangle_{\Lambda(R,x^0)} (A) =
$$
$$
e^{-{q^2 \over 
2\beta_\chi}((\delta _x -\delta_{R_+}),\Delta_\Lambda^{-1} (\delta_x - 
\delta_{R_+}))}
e^{i(E_\Lambda (x, R_+), A)} F ( E_\Lambda (x, R_+)| {2\pi\over q}  d A).
\eqno(4.30)
$$

From (4.26) and (4.30) it follows that $d \mu (j_x)$ behaves, for large 
$\beta_\chi$, as an approximate Dirac $\delta$ measure around 
$E_\Lambda (x, R_+)$.
This, in turn, implies that $d\Sigma (j_x,j_y; \omega)$ is peaked around

$$
B_\Lambda (x, R_+) + B_\Lambda (R_-, y) +^*\omega + 2 \pi ^*\delta_{<R_+ R_->}, 
\quad {\rm with} \quad  B_\Lambda =^* E_\Lambda
$$

Here we see the dual version of the phenomenon occurring for charged fields: 
the magnetic current appearing in the definition of monopole Green functions is
forced to be $2\pi {\bf Z}$ valued so as to respect Dirac quantization 
condition. But its 
average, at sufficiently large scales, approximates the ordinary magnetic field
$B$, of the classical monopole.

The deviation from a $\delta$--measure of $d \mu_{R_+} (j_x)$ is 
``small", for $\beta_\chi$ large. This may explain why, in numerical 
simulations [33], a naive application of the definition of monopole Green 
functions in theories without 
matter fields appears to give reasonable results.

Redefining $\beta_\chi = q^2 \tilde\beta_\chi$, keeping $\tilde\beta_\chi$ 
fixed and taking the limits $q \nearrow \infty, \kappa \searrow 0$ one 
recovers the monopole Green functions discussed in sect. 3.3.

Using a Peierls type argument and renormalization group techniques discussed 
in [2,3,12,24], we may argue that  
the monopole Green functions in a theory with matter fields have the 
same qualitative large 
distance behaviour as the monopole Green functions 
in a pure $U(1)$ gauge theory, if $\beta$ and $\kappa^{-1}$ or 
$q^2/{\beta}$ are sufficiently large. (In the dual model the renormalization 
of the electric currents $\rho$ requires $\beta$ to be large, and the 
renormalization of the monopole currents $d \gamma$ requires 
$\kappa^{-1}+O(q^2/{\beta})$ to be large.)

\vskip 0.3truecm
{\sl 4.5 \ Dyons}
\vskip 0.3truecm

Model B in Villain form is self--dual in the limit $\kappa \nearrow \infty$, 
where it becomes a ${\bf Z}_q$ gauge theory. If a topological term (2.9) is 
added one can show, see [11], that a modified duality transformation is an 
approximate symmetry. When combined with the 2$\pi$--shift of $\Theta$ it 
generates an $SL (2,{\bf Z})$ duality group of approximate (large--scale)
symmetries of the model.

Here we briefly describe how to derive the duality symmetry and then discuss 
its action on charged and  monopole Green functions. 
Green functions of dyon fields are obtained in a natural way.

A wedge 
product on the lattice compatible with O.S. positivity can be defined [17]
as follows: Label a (positively oriented) $k$-cell in the lattice by a site,
$x$, and a set of $k$ directions $\underline{\mu}=\{\mu_1, ... 
\mu_k \}$; i.e.,

$$
c_k = (x; \mu_1,..., \mu_k) = x + \sum^k_{i=1} \xi^{\mu_i} 
e_{\mu_i}, \eqno(4.31)
$$

where $\xi^{\mu_i} \in [0, 1]$ and $e_{\mu_i}$ is the unit vector in the
$\mu_i$ direction.

Given a $p$--form $A$ and a $q$--form $B$ we define

$$
A \wedge B (x; {\underline{\mu}})=
\sum_{\mu_j \in {\underline{\mu}}} 
\epsilon_{{\mu_1} ...\mu_{p+q}} {1\over 2} \{A (x; \mu_1, ..., \mu_p)
B (x+ e_{\mu_1} + ... + e_{\mu_p}; \mu_{p+1}, ..., \mu_{p+q}) 
$$
$$
+{1\over 2} A (x; r \mu_1, ..., r\mu_p)
B (x + e_{r\mu_1} + ... + e_{r\mu p}; r \mu_{p+1}, ..., r\mu_{p+q}) \} 
\eqno(4.32)
$$

where $r$ inverts the time direction, leaving the other ones unchanged.

It is easy to verify that

$$
d(A\wedge B) = d A\wedge B + (-1)^p A \wedge dB \eqno(4.33)
$$

but in general

$$
A\wedge B \not = - (-)^{p+q} B\wedge A
$$

Furthermore, on the lattice, we do not have an analogue of the continuum 
identity

$$
\int A\wedge B = (A, ^*B) = (^*A,B) \eqno(4.34)
$$

for $p+q=4$, where $^*$ is the Hodge dual and (,) denotes the inner product on
forms in $\bf{R}^4$.

From the definition (4.32) it follows, however, that we can define two 
modified star operations, $A \mapsto ^\star A, A \mapsto A^\star$, 
such that for $p+q=4$,

$$
\sum_{c_4} (A\wedge B) (c_4) = (A, ^\star B) = (A^\star,B)\eqno(4.35)
$$

and with the property that if $\delta^*A=0$, then $\delta^\star A=0=\delta 
A^\star$, but $A^\star\not= ^\star A$ and $A^{\star \star} \not = A, 
^{\star \star}A \not =A$. Furthemore, if 
$A$ is ${\bf Z}$--valued, $A^\star$ and $^\star A$ are ${\bf Z}/2$--valued.

Using the rule (4.33) one can rewrite the topological term (2.9) in terms of 
the magnetic currents $m=dn:$ up to boundary terms vanishing if  
$0$--Dirichlet b.c. are imposed on $d\theta$ and $n$, it can be rewritten as

$$
i {\Theta\over 4\pi^2} q \sum_{c_4} (m\wedge \theta + \theta \wedge m + 
n\wedge n) (c_4) \eqno(4.36)
$$

Its meaning is clear: the first two terms associate to every magnetic 
current $m$ a line of total electric flux ${\Theta\over 2\pi} q$ along the 
support of $m^\star$ and $^\star m$ (Witten effect [34]); the last term is the
self--intersection form of the surface $^*n$.

To discuss the duality group it is convenient to define 

$$
A_\pm = {A \pm A^\star\over 2}
$$
 
for each $k$--form $A$, and to introduce, following [11], 
the complex coupling constant parameter

$$
\zeta = - i {2\pi \beta \over q} + {\Theta \over 2\pi} \eqno(4.37)
$$

(and  its complex conjugate, $\bar\zeta$) .

The pure gauge action of the model can be rewritten as

$$
S_0 (\theta, n) = {iq\over 2\pi} \zeta ||(d\theta + n)_+ ||^2 - {iq\over 
2\pi} \bar\zeta ||(d\theta+ n)_-||^2 \eqno(4.38)
$$

and, integrating over the matter field, the partition function is given by

$$
Z = \sum_n \int \prod_{<ij>} d\theta_{<ij>} e^{-S_0 (\theta,n)} 
\sum_{\rho:\delta\rho=0} e^{iq(\theta,\rho)} \eqno(4.39)
$$

The $SL(2,{\bf Z})$ group acts on $\zeta$ by 

$$
\zeta\rightarrow \zeta^\prime = {A\zeta + B \over C\zeta + D}, \quad AD-BC \not 
=0
$$

Its generators are called $S: \zeta\rightarrow \zeta^{-1}$, and $T: 
\zeta\rightarrow \zeta+1$.

We now show why $S$ is an approximate symmetry of the partition function.

Introducing real 2--forms $\lambda, \bar\lambda$, denoting by $[\sigma]$
the equivalence class of $\bf{Z}$--valued two--forms, i.e., $[\sigma]= 
\{\sigma^\prime - \sigma = \delta V, V (c_3) \in {\bf Z} \}$, and 
performing the change of variables $\{\theta, n\}\rightarrow \{A, m\}$, as 
in previous section, one obtains that

$$
Z = \sum_{m:dm=0} \sum_{[\sigma]} \int \prod_{<ij>} dA_{<ij>} \prod_p 
d\lambda_p \prod_p d\bar\lambda_p
$$
$$
e^{-{i\over 2\pi q\zeta} ||\lambda||^2} e^{+{i\over 
2\pi q\bar\zeta}||\bar\lambda||^2}
e^{i(dA+n[m], {(\lambda)_+\over 2\pi} +{(\bar\lambda)_-\over 2\pi}+ q\sigma)}
$$

Integrating out $A$ we obtain the constraint

$$
\delta ({(\lambda)_+\over 2\pi} + {(\bar\lambda)_-\over 2\pi} + q\sigma) = 0
\eqno(4.40)
$$

If the identity $A^{\star \star}=A$ were true then it 
would follow from (4.40) that 
there exists a real--valued 3--form, $\xi$, such that

$$
\lambda = (\delta \xi - 2\pi q\sigma)_+\quad,\qquad
\bar\lambda = (\delta\xi - 2\pi q \sigma)_-
$$

Formally, $A^\star$ and $A^*$ have the same continuum limit. On large scales,
we may therefore replace $^\star$ by $^*$. Then $A_+$ and $A_-$ are just the 
selfdual and anti--selfdual components of $A$ and, indeed, $A^{**} =A$.

So, within this approximation,

$$
Z \sim \sum_{m:dm=0} \sum_{[\sigma]} \int \prod_c d\xi_c
e^{-i {1\over 2\pi q\zeta} ||(\delta\xi -2\pi q\sigma)_+||^2} e^{i{1\over 
2\pi q \bar\zeta} ||(\delta \xi - 2\pi q \sigma)_-||^2}
e^{i({n[m]\over 2\pi}, \delta\xi)}. \eqno(4.41)
$$

Passing to the dual lattice and introducing a $U(1)$--valued 1--form 
$\tilde\theta$ and a ${\bf Z}$--valued 1--form $\ell$ and  setting

$$
\xi = q^*\tilde\theta +  2\pi q^* \ell,
$$

we can rewrite (4.41) as

$$
Z \sim \sum_{m:dm=0} \sum_\sigma \int \prod_{<ij>} d\tilde\theta_{<ij>} 
e^{-{iq\over 2\pi\zeta} ||(d\tilde\theta- 2\pi^* \sigma)_+ ||^2}
e^{+{iq\over 2\pi \zeta} ||(d\tilde\theta - 2\pi^*\sigma)_- ||^2}  
e^{iq({^*m\over 2\pi}, \tilde\theta)} \eqno(4.42)
$$

Equation (4.42) proves that the $S$ generator $SL (2, {\bf Z})$ induces an
approximate symmetry exchanging

$$
\zeta \rightarrow {1 \over \zeta} \qquad m \rightarrow -2\pi ^*\rho
\qquad \rho \rightarrow {^*m\over 2\pi}.
$$

Analogously, the $T$--generator induces an approximate symmetry exchanging

$$
\zeta \rightarrow \zeta +1 \qquad n \rightarrow n 
\qquad \rho\rightarrow \rho - {m^\star\over 2\pi} - 
{^\star m\over 2\pi}
$$

as one can easily see from (4.15) (4.36); the symmetry becomes exact for 
$T^2$ since ${m^\star\over \pi}$ and ${^\star m \over \pi}$ are actually 
${\bf Z}$--valued, as the original $\rho$ current.

Our construction of the charged 2--point function, $G(x,y)$, is based on 
the introduction of a weighted sum of Wilson loops

$$
W_\omega (j_x,j_y)= e^{i q(\theta, j_x + j_y + ^*\omega +
\delta_{ <R_+ R_->})}\eqno(4.43)
$$

where $\omega$ is a current line from $x$ to $y$,
and our construction of the monopole 2--point function $G^*(x,y)$ is based
on the introduction of a weighted sum of 't Hooft loops

$$
D_\omega (j_x,j_y)=
$$
$$
e^{-i {q \zeta \over 2\pi} \{||(d\theta+ n + \Sigma (j_x, j_y; \omega))_+||^2 
- ||(d\theta -n)_+ ||^2\}}
e^{i{q \bar\zeta \over 2\pi} \{||(d\theta +n+\Sigma (j_x, j_y; \omega))_-||^2 - 
||(d\theta +n)_-||^2\}} \eqno(4.44)
$$

with the notations of previous sections.

The above discussion makes it clear that the $S$ generator of $SL (2,{\bf Z})$
approximately exchanges the two correlation functions in the dual 
models, while the $T$ generator acts (approximately) by multiplying the 
't Hooft 
loop (4.44) by the Wilson loop (4.43), raised to the power $-2$, hence 
producing the 2--point function of a dyon, 
whose electric charge is $-2q$ at $\Theta =0$. 

\vskip 0.3truecm
{\bf 5. \ 't Hooft--Polyakov monopoles}

\vskip 0.3truecm
{\sl 5.1 Preliminaries}
\vskip 0.3truecm

In order to explain our proposal for the  construction  of Green  functions
for the quantum 't Hooft--Polyakov monopole,
it is useful to recall the definition of 
topological invariants characterizing the classical 't Hooft--Polyakov 
monopole in the Georgi--Glashow model and then to exhibit their 
lattice counterparts.

The field content of the classical Georgi--Glashow model consists of 
a connection $A$ on a 
principal $SU(2)$--bundle, $P_{SU(2)}$, on ${\bf R}^4$  and a scalar field 
$\Phi$ with values in $su(2) = Lie (SU(2))$, which is a section of a vector 
bundle associated to $P_{SU(2)}$ with fiber $su(2)$ equipped with the 
adjoint action of $SU(2)$.

The Lagrangian density of the model is given by

$$
{\cal L} = {1\over 2} |F_A|^2 + {1\over 2} |D_A \Phi|^2 - {\lambda\over 8} 
(|\Phi|^2 -1)^2 \eqno(5.1)
$$

where $F_A$ is the curvature of $A$, and $D_A$ is the covariant derivative

$$
D_A \Phi = d\Phi + [A, \Phi]
$$

Moreover $| \cdot |$ denotes the Killing norm on $su(2)$. 
Let us assume that $A$ and $\Phi$ verify appropriate regularity 
conditions, and $|1 - | \Phi (\vec x)||, |D_A 
\Phi| (\vec x), F_A (\vec x)$ have 
decay properties at $\infty$ discussed in detail in [35]. Then, at fixed 
time, a finite--energy static configuration 
$(A, \Phi)$, with $A_0 =0$, defines a homotopy class,
$[(A, \Phi)] \in \pi_2 (SU (2)/ U (1)) \sim {\bf Z}$, labelled by an integer
topological charge

$$
N= {1\over 4\pi}  \int_{{\bf R}^3} {\rm Tr}(F_A \wedge D_A \Phi) 
= {1\over 4\pi} \int_{S^2_\infty} {\rm Tr }(\tilde\Phi F_A) \eqno(5.2)
$$

where $S^2_\infty$ is the 2--sphere at infinity. Furthermore, the field $\Phi$ 
defines a homotopy class $[\Phi] \in \pi_2 (S^2)$ which can be 
characterized by the Kronecker index

$$
N= {1\over 4\pi} \int_{S^2_R} {\rm Tr} (\tilde\Phi \wedge d\tilde\Phi \wedge d 
\tilde\Phi); \eqno(5.3)
$$

with $\tilde\Phi = \Phi/|\Phi|$, and $S_R = \{|x|= R\}$ provided $R$ is 
sufficiently large. Then (5.3) is independent of $R$, and $[\Phi] = [(A,\Phi)]$.
A solution of the equations 
of motion with $N\not = 0$ is called a classical 't Hooft--Polyakov 
monopole. 

The magnetic charge of a Dirac monopole in ${\bf R}^3$ is given 
by the first Chern number of a $U(1)$--bundle over a topological 2--sphere 
containing the monopole.[This geometrical definition differs from the 
one used previously by a factor $1/(2 \pi)$.]
We show how one can associate to a configuration $(A, \Phi)$ and a cube $c$
in ${\bf R}^3$ a $U(1)$--bundle in such a way that the first Chern number 
of the bundle can be identified with the magnetic charge contained in $c$. 
We then relate the magnetic charge to the topological charge previously 
defined. 

We choose a cube $c$ in ${\bf R}^3$; its boundary, $\partial c$, is 
homeomorphic to a 2--sphere. The restriction of an $SU(2)$ connection $A$ 
to $\partial c$ can be viewed as a connection $A|_{\partial c}$ on an 
$SO(3)$ bundle over $\partial c$. There is then an $SO(3)$ gauge 
transformation mapping $(A, \Phi)|_{\partial c}$ to $(\bar A, 
|\Phi|\sigma_3)|_{\partial c}$.
Projecting $\bar A$ to a Cartan subalgebra of $su(2)$ one
obtaines a $U(1)$ connection, $a$. This projection is called ``abelian 
projection". It has been introduced by 't Hooft [36] to express Yang--Mills 
theories in terms of abelian gauge fields, charges and monopoles.
We identify the magnetic charge contained inside $c$ with the first Chern 
number of the curvature, $F_a$, of $a$.

The relation between $\Phi, \bar A$ and $a$ can be described in terms of 
the homotopy exact sequence

$$
0 \sim \pi_2 (SO(3)) \rightarrow \pi_2(S^2) 
\mathop{\longrightarrow}\limits^\partial
\pi_1(U(1)) \mathop{\longrightarrow}\limits^{i_*}
\pi_1(SO(3)) \rightarrow \pi_1(S^2) \sim 0
$$

where

$$ 
\partial: n \in {\bf Z} \sim \pi_2(S^2) \rightarrow 2 n \in {\bf Z} \sim 
\pi_1(U(1))
$$

and

$$
i_* : n \in {\bf Z} \sim \pi_1(U(1)) \rightarrow n {\rm mod} \ 2 \in
{\bf Z}_2 \sim \pi_1(SO(3))
$$

The group $\pi_1 (U(1))$ classifies $U(1)$--bundles over
the (topological) 
2--sphere $S^2$, here identified with $\partial c$. The integer 
$n\in \pi_1 (U(1))$ is given in terms of a connection $a$ on a principal 
$U(1)$--bundle over $\partial c$ by

$$
n = {1\over 2\pi} \sum_{p\in\partial c} \int_p F_a \eqno(5.4)
$$

where $p$ denotes faces of the cube $c$.

The group $\pi_1 (SO(3))$ classifies $SO(3)$--bundles over $S^2 \simeq 
\partial c$.  We have the following relation between $n$ mod 2 $\in \pi_1 
(SO(3))$ and a connection  $\bar A$ on a principal $SO(3)$ bundle over 
$\partial c$:

$$
e^{i\pi n} = {\rm exp} [i \{{\rm arg} \sum_{p\in\partial c} Tr P 
(e^{i\oint_{\partial p}
\bar A^{(p)}}) \}], \eqno(5.5)
$$

where $\bar A^{(p)}$ is the connection 1--form of an $SU(2)$ bundle over 
$p$ obtained by lifting the $SO(3)$--bundle over $p$.
If $e^{i\pi n} \not =1$ one cannot extend the $SO(3)$--bundle to the interior 
of the cube $c$. This signals the presence of an odd number of singular 
${\bf Z}_2$--monopoles of $SO(3)$ inside $c$.

If $\tilde\Phi$ is a map from $S^2_R$ to $S^2$ of Kronecker index $N$ 
then the first Chern number corresponding to
$\partial [\tilde\Phi]$ is 
$2N$. Hence 't Hooft--Polyakov monopoles have even magnetic charge and,
with an abuse of language, following [37], the term 't Hooft--Polyakov 
monopole stands for an arbitrary monopole  configuration with   
even magnetic charge.
The most general configuration inside a cube $c$ in ${\bf R}^3$ is a 
combination of 't Hooft--Polyakov monopoles and ${\bf Z}_2$ monopoles 
characterized by odd magnetic charge.

Next we explain how to define the magnetic charge on the
lattice by means of the abelian projection, following [38].

In model C, we start  by choosing a unitary gauge by means of an $SO(3)$ 
gauge transformation $W$. In this gauge the fields are given by

$$
\bar\Phi_i = W_i \Phi_i W^{-1}_i = \sigma_3
$$
$$
\bar g_{<ij>} = W_i g_{<ij>} W^{-1}_j \eqno(5.6)
$$

We write the $SU(2)$--valued gauge field $\bar g_{<ij>}$ as a product

$$
\bar g_{<ij>} = C_{<ij>} u_{<ij>} (\theta) 
$$

with

$$
C_{<ij>}= \left(\matrix{(1-|c_{<ij>}|^2)^{1\over 2} & - c^*_{<ij>} \cr
c_{<ij>} & (1 - |c_{<ij>}|^2)^{1\over 2} \cr} \right),
$$
$$
u_{<ij>} (\theta) = e^{i{\theta_{<ij>} \over 2} \sigma_3}, \eqno(5.7)
$$

where $c_{<ij>} \in {\bf C}, c^*_{<ij>}$ denotes its complex 
conjugate and ${\theta_{<ij>}\over 2} = \arg (\bar g_{<ij>})_{11}$.

Under a $U(1)$--gauge transformation $\{e^{-i\Lambda_j {\sigma_3\over
2}}\}$ of the original fields, $\theta$ and $c$ transform as

$$
\theta \rightarrow \theta + d\Lambda \quad, \qquad
c_{<ij>} \rightarrow c_{<ij>} e^{{i \over 2}\Lambda_i}
e^{{i \over 2}\Lambda_j} 
$$

Hence $\theta$ is a $U(1)$--gauge field, and $c$ is a charged field 
of charge 1.

We define the magnetic charge in a lattice cube $c$ by

$$
m_c (\theta) = {1\over 2\pi} \sum_{p\in \partial c} (d\theta)_p, 
\eqno(5.8)
$$

where $d\theta$ is restricted to the range $[-2\pi, 2\pi]$.
Equation (5.8) is the lattice analogue of eq. (5.4). A ${\bf Z}_2$-- charge
in a cube $c$ is defined by

$$
e^{i\pi z_c (g)} = e^{i \sum_{p\in \partial c} {\rm arg} \chi(g_{\partial 
p})}.
\eqno(5.9)
$$

A plaquette $p$ where

$$
e^{i{\rm arg}\chi(g_{\partial p})} = -1
$$

can be identified as the location of a Dirac string of a 
${\bf Z}_2$--monopole intersecting the plane containing $p$.

The relation established between (5.4) and (5.5) can be translated to 
the lattice as

$$
e^{i\pi z_c (g)} = e^{i\pi m_c (\theta)}. \eqno(5.10)
$$

With the help of the abelian projection, the phase transitions in model C 
have been analyzed numerically [39] in terms of condensations of 
magnetic currents.

\vskip 0.3truecm
{\sl 5.2  Monopole Green functions} 
\vskip 0.3truecm

In model C, the magnetic charge of 't Hooft--Polyakov monopoles is even and
the electric charge of the matter field is integer. Hence we are facing a 
situation analogous to the one discussed in sect 4.3, for $q=2$, in models 
of class B. Following the ideas in that section, we propose to construct 
the two--point 
function of a monopole of magnetic charge 2 by summing over pairs of 
fluctuating strings of magnetic charge 1 with end points at the 
location of the monopole.

As discussed in the previous section, these strings can be identified as 
Dirac strings of ${\bf Z}_2$--monopoles. They can be introduced by means of
't Hooft disorder fields. For a 2--surface $\Sigma$ bounded by a lattice loop
$L$, the disorder field is defined by

$$
D(\Sigma)=e^{-[S_0 (g e^{i\Sigma \sigma_3}) - S_0 (g)]}\eqno(5.11)
$$

where $S_0$ is given in (2.8), $\Sigma$ is a 2--form with values in 
$\{0, \pi\}$, whose support is dual to the surface $\Sigma$.

The expectation value of $D(\Sigma)$ depends only on $L$. Clearly
$e^{i \Sigma \sigma_3}$ takes values in $\{0,1\} \sim {\bf Z}_2$,
which is the center of the gauge group $SU(2)$ .

In order to construct the monopole two--point function we start, by 
introducing    
a $U(1)$ scalar field $\chi$ on the sublattice 
$\Lambda (R,t)$. The field $\chi$ is minimally coupled to a real gauge 
field $A$ with coupling constant
$\pi$. The action for $\chi$ is given by $S_\Lambda (\chi,r)$, a functional 
defined in (4.25), with $q=2$.

Since $S_\Lambda (\chi, r)$ is periodic in $A$ with period 2, 
$\langle e^{i \chi_x} e^{-i \chi_{R+}} \rangle_{\Lambda(R,x^0)} (A)$
is the Fourier transform of a complex 
measure $d \mu_{R_+} (j_x)$ on the space of currents $j_x$ with values 
in $\pi{\bf Z}$ which are
constrained by $\delta j_x=2 \pi (\delta_x - \delta_{R_+})$; see formula 
(4.26). We propose to define the two--point function for a 't Hooft--Polyakov 
monopole of magnetic charge 2 by

$$
G^* (x,y) = {\lim\limits_{R\rightarrow \infty}} c_{x,1}^R c_{y,1}^R 
\langle \int d \mu_{R_+} (j_x) d \mu_{R_-} (j_y) 
D_\omega (j_x,j_y) \rangle 
\eqno(5.12)
$$

with $\Sigma$ constrained by

$$
^*d \Sigma(j_x,j_y; \omega) = j_x + j_y + \omega + 2\pi \delta_{<R_+R_->}, 
\eqno(5.13) 
$$

where $\omega$ is a 3--form taking values in  $\{0, 2\pi\}$ whose 
support is dual to a path connecting $x$ to $y$.

As the definition (5.12) involves a 't Hooft disorder field, the expectation
value on the r.h.s. is independent of the choice of $\Sigma$, given 
$\omega$,
and, since $\omega$ takes value in $\{0, 2\pi\}$ , also of $\omega$.

In Appendix B it is shown, using the abelian projection, that the disorder
field $D_\omega (j_x, j_y)$ introduces a source for 't Hooft--Polyakov 
monopole currents. For this purpose, a representation of $<D_\omega (j_x, j_y)>$
as a sum over configurations of magnetic currents of even 
charge, with boundary given by $\{x, y\}$, is derived. 
This representation makes 
precise the idea that, in the abelian projection, the effect of the 
disorder field is to introduce a shift of $d\theta$ by $2 \Sigma (j_x, j_y; 
\omega)$. Hence, according to definition (5.8), the magnetic 
charge in a cube $c$ is changed by $2 (d\Sigma)_c$.

\smallskip
\underbar{Remark 5.1} In (5.10) one may replace the matrix $\sigma_3$ by 
$\Phi_x$ at the plaquette $p=(x; \mu_1, \mu_2)$, since, in the abelian 
projection, $\Phi_x$ is replaced by $\sigma_3$, and the computation of the 
magnetic charge is then unchanged.

\smallskip
\underbar{Remark 5.2} The above construction cannot be adapted to obtain Green 
functions for the ${\bf Z}_2$--monopoles. For such Green functions, 
the current $\omega$ would take values in $\{0,\pi\}$, and the 
expectation value of the disorder field $D_\omega (j_x, j_y)$ would 
depend on $\omega$, hence on the Dirac strings of the ${\bf
Z}_2$--monopoles, which is unphysical (see Appendix B).
\smallskip 

One can argue that, in the Coulomb phase of model C correlation functions 
of non--zero total magnetic charge vanish (as proved at 
$\beta_H=\infty$, where the model reduces to a $U(1)$ theory), so 
that we expect the appearence of superselection sectors 
labelled by an even magnetic charge . According to the previous 
section, one may identify these sectors as 't Hooft--Polyakov monopole 
sectors. In the confinement phase of the model, Green 
functions of non--zero total charge are expected not to vanish due 
to ``monopole 
condensation" (this is proved at $\beta_H=\infty$). 
The large distance behaviour of the two--point monopole Green
function could then be used to detect phase transitions in the 
Georgi--Glashow model, as in the $U(1)$ gauge theory.

\vskip 0.3truecm
{\sl 5.3 Continuum}
\vskip 0.3truecm

In the final section we sketch how to define Green functions for quantum 
't Hooft--Polyakov monopole fields in the formal continuum limit of the 
Georgi--Glashow model.

It turns out that it is more convenient to work in a first--order 
formalism for the Yang--Mills field. (A formal relation of this formalism 
to ordinary Yang Mills and BF theories is discussed in [40]).

We introduce a 2--form $B$ with values in $su(2)$, which is a section 
of the vector bundle of 2--forms associated to $P_{SU(2)}$. 
We replace the Yang--Mills euclidean action

$$
S_0 (A) = {1\over 2} \int |F_A|^2 d^4 x
$$

by

$$
S_0 (A,B) = i \int {\rm Tr} (F_A \wedge B) + {1\over 2} \int |B|^2 d^4 x 
\eqno(5.14)
$$

In the functional integral approach, integrating over $B$ with the 
(white--noise) gaussian measure corresponding to the second term in (5.14),
formally yields the standard weighting factor, $e^{-S_0(A)}$, for 
Yang--Mills fields.

In the first order formalism, we use 

$$
D_\omega (j_x, j_y)= [e^{i\int {\rm Tr} B \wedge \tilde\Phi \Sigma (j_xj_y; 
\omega)}]_{ren} \eqno(5.15)
$$

as our 't Hooft disorder field,
with the notations of the previous section adapted to the continuum. The 
notation $[\cdot]_{ren}$ indicates that an ultraviolet multiplicative 
renormalization is necessary. 
For the lattice theory, (5.15) corresponds to the choice mentioned in 
Remark 5.1.

\vskip 0.3truecm
We wish to comment on the relation between our construction of monopole 
Green functions, based on expression (5.15) for the disorder operator, and 
the conventional semi--classical analysis of 't Hooft--Polyakov monopoles. 
For this purpose, we consider a monopole two--point function with $y= rx$ 
(where, as usual, $r$ denotes time reflection), and $x= (x^0, \vec 0)$. We 
are interested in the asymptotic behaviour of this Green function, as $x^0$
becomes large. One strategy to analyze this behaviour is to attempt to 
evaluate the functional integral defining the Green function with the help 
of semi--classical techniques. In a semi--classical approximation, an 
evaluation of the expectation value

$$
{\lim\limits_{R\rightarrow \infty}} \langle \int d\mu_{R_+} (j_x) d\mu_{R_-}
(j_y) D_\omega (j_x, j_y) \rangle
$$

of the disorder operator in the formal functional 
measure of the Georgi--Glashow model is accomplished 
through an expansion of the functional measure around a back--ground field 
configuration. In the unitary gauge of the abelian projection, the disorder
operator creates a ``mean background gauge field" with curvature $F_{\bar A}
(z)$ approximately given by

$$
\sigma_3 \vec B (\vec z)
$$

for $z= (z^0, \vec z)$, with $|z^0| << x^0$, where $\vec B (\vec z)$ is the
rotation--covariant, static magnetic field generated by a magnetic monopole
of magnetic charge 2, located at the origin.

If, instead of the unitary gauge, we choose a gauge with the property that

$$
\tilde\Phi (z) = {\vec z \cdot \vec \sigma \over |\vec z|} \eqno(5.16)
$$

the mean background gauge field has a field strength $F_{\bar A} (z)$ 
approximately given by

$$
\vec B^{H-P} (\vec z) = {\vec z \cdot \vec \sigma \over |\vec z|} 
\eqno(5.17)
$$

for $z= (z^0, \vec z)$, with $|z^0| << x^0, |\vec z|$ large.

The field in (5.17) describes the large--distance behaviour of the field 
strength of the classical 't Hooft--Polyakov monopole solution [5].

We should comment on the notion of ``mean background (gauge) field" used in
the arguments above: Every field configuration contributing to the 
expectation value of a disorder field $D_\omega (j_x, j_y)$ actually 
satisfies the non--abelian version of Dirac's quantization condition 
[28,37]. Such a field configuration is therefore singular near the support of 
the magnetic flux currents, $j_x$ and $j_y$.
However, after integrating over $j_x$ and $j_y$ with the
complex measure 
$d\mu_{R_+} (j_x) d\mu_{R_-} (j_y)$ and taking the limit $R\rightarrow 
\infty$, and after integrating out the high--frequency (short--distance) 
modes of the fields, i.e., after ultraviolet renormalization (``coarse 
graining"), the resulting background field configuration approaches the one
described above. This is because the constraint of flux quantization of the 
Dirac strings is softened under renormalization and is actually expected to
scale to zero in the limit of very large distances scales. This phenomenon 
is analogous to the one discussed in the abelian models of class B.

Taking it for granted that, on large distance  scales, the background field
configuration described in (5.16), (5.17) dominates the functional integral
appearing in the (numerator of the) monopole Green function $G(x, r x)$, as
$x^0$ becomes large, the relation of our approach to the conventional 
semi--classical analysis of (e.g., the mass of) the 't Hooft--Polyakov 
monopole becomes clear.

We conclude this section with a comment on dyons. If a topological term 

$$
i {\Theta\over 16 \pi^2} \int Tr (F_A \wedge F_A)
$$

is added to the action of the Georgi-- Glashow model then the excitation 
spectrum of the model contains \underbar{dyons}. 

Dyon Green functions can be constructed with the help of disorder 
fields. The construction is analogous to the one discussed in the 
context of the Cardy--Rabinovici model, in sect. 4.5. If $\Theta$ is not an
integer multiple of $2\pi$ dyons carry a fractional charge. As $\Theta$ 
approaches an integer multiple of $2\pi$, the dyons correspond to those 
first described by Julia and Zee [41]. The spectrum of the theory is 
periodic in $\Theta$, with period $2\pi$.

\vskip 0.3truecm
{\bf Appendix A} (Gaussian evaluation of $F(E|dA)$) 
\vskip 0.3truecm

[In this appendix all symbols referring explicitly to the sublattice 
$\Lambda(R,t)$ are omitted, e.g., we write $d$ instead of $d_\Lambda$ etc..].

To analyse $F(E |d A)$ it is convenient to work on the dual lattice and 
introduce an auxiliary gauge field, $C$: 

$$
F(E|d A) = {\int \prod_{<ij>} dC_{<ij>} e^{-{1\over 2\beta_\chi} ||dC||^2}
\sum_{\rho:\delta\rho=0} e^{-i(C,^*dA)} e^{i(dC-^*E,\sigma(\rho))}\over 
\int\prod_{<ij>} dC_{<ij>} e^{-{1\over2\beta_\chi} ||dC||^2} 
\sum_{\rho:\delta\rho=0} e^{-i(C,^*dA)} e^{i(dC,\sigma(\rho))}}
\eqno({\rm A}.1)
$$

where, comparing with (4.18), $\rho=^*v$ and $\sigma(\rho)$ is a fixed, 
integer--valued solution of 

$$
\delta\sigma(\rho)=\rho.
$$

Following the techniques developed in [12] (see also [2,3]), we divide a
configuration of currents, $\rho$, into (not 
necessarily connected) networks, $\{\rho_\alpha\}$, such that the 
distance
$d(\rho_\alpha, \rho_\beta) \geq 2$, $\alpha \not=\beta$.

We now renormalize the activities of the current networks, 
$\{\rho_\alpha\}$, using the method of complex translations.

Let ${\cal B} (\rho_\alpha)$ be a set of links in the support of 
$\rho_\alpha$ such that
two links in ${\cal B}(\rho_\alpha)$ do not belong to a common plaquette 
and such that

$$
\sum_{{<ij>}\in{\cal B}(\rho_\alpha)} |\rho_{\alpha <ij>}|^2 \geq c 
(\rho_\alpha,\rho_\alpha),
$$

where 

$c^{-1}=$ card $\{\ell^\prime: \ell^\prime \not = \ell, \{\ell, \ell^\prime\}
\in \partial p$ for some $p\}=18$.

We set

$$
\rho_\alpha^\prime =\rho_\alpha|_{{\cal B}(\rho_\alpha)} \quad 
\rho^\prime=\sum_\alpha \rho_\alpha^\prime
$$

and perform the complex translation

$$
C\rightarrow C + i {\beta_\chi \over n} \rho^\prime  + i \beta_\chi
\Delta^{-1}{}^*dA
$$

where

$$
n = {\rm card} \{p:\partial p \in <ij>\}=6
$$

As a result every current network $\rho_\alpha$  obtains a complex activity

$$
z_A (\rho_\alpha, C, E) = e^{-{\beta_\chi\over 2} [||\rho^\prime_\alpha||^2 
+2 (\rho_\alpha, \Delta^{-1}{}^*dA)]} e^{i(d\bar C - ^*E, \sigma(\rho))}
$$

in the numerator of (A.1), and an activity $z_A (\rho_\alpha; C) \equiv z_A 
(\rho_\alpha, C,0)$ in the denominator of (A.1), where $\bar C = C - {1\over n} 
(\delta dC)|_{\rho^\prime_\alpha}$.

If $dA$ is sufficiently small so that 

$$
|z_A (\rho_\alpha, C)|, |z_A (\rho_\alpha, C, E)| \leq e^{-c \beta_\chi 
||\rho_\alpha||^2} \eqno({\rm A}.2)
$$

the current networks form a dilute gas, and we can exponentiate their 
contributions in the form of a Mayer series:

$$
F(E|dA) = {\int \prod_{<ij>} dC_{<ij>} e^{-{1\over 2 \beta_\chi} ||dC||^2 + 
\sum_{\cal L}a_T ({\cal L})  z_A ({\cal L},C,E)}
\over \int \prod_{<ij>} 
dC_{<ij>} e^{-{1\over\beta_\chi} ||dC||^2 + \sum_{\cal L}a_T({\cal L})
z_A({\cal L}, C)}} 
\eqno({\rm A}.3)
$$

where ${\cal L}$ is a collection of current networks $\rho$ in which
a single $\rho$ can occur an arbitrary number of times ,
$a_T ({\cal L})$ is a standard combinatorial factor which enforces all 
${\cal L}$ to have connected supports and

$$
z_A ({\cal L},C,E)=\prod_{\rho \in {\cal L}} z_A (\rho,C,E).
$$

We give an approximate evaluation of (A.3) by keeping only the leading 
contribution in ${\cal L}$, which correspond to plaquette terms, and 
expanding the activities $z_A$ to  second order in $C, A,E$.

$F(E|dA)$ is then approximated by

$$
{\int \prod_{<ij>} dC_{<ij>} e^{\{-{1\over 2\beta_\chi} ||dC||^2 - 
\sum_p e^{-O(\beta_\chi)} (1+{\beta^2\over 2} (^*A^T)^2_p) (d\bar C-^*E)^2_p +
i \beta_\chi (^*A^T)_p (d\bar C)_p]\}} \over \int \prod_{<ij>} dC_{<ij>} 
e^{\{-{1\over 2\beta_\chi} ||dC||^2 - \sum_p e^{-O(\beta_\chi)} 
(1+{\beta^2_\chi\over 2} (^*A^T)^2_p) (d\bar C)^2_p + i \beta_\chi 
(^*A^T)_p 
(d\bar C)_p]\}}}
\eqno({\rm A}.4)
$$

where $A^T= \delta \Delta^{-1} dA$, and we have used that

$$
(d\bar C,^*E)=0.
$$

Explicit evaluation of (A.4) to second order in $A$ gives

$$
F^G (E|dA)={\rm exp} [-e^{-O(\beta_\chi)} ({1\over 2} ||E||^2 + 
{\beta^2_\chi\over 2} ||A^T \cdot E||^2)]
$$

\vskip 0.3truecm
{\bf Appendix B}
\vskip 0.3truecm

In this appendix we exhibit a representation of $\langle D_\omega (j_x, 
j_y)\rangle$ in model C in terms of  
magnetic currents. We start by replacing the $SU(2)$ gauge field $g$ with a
couple of new variables $\{U,\sigma\}$. $U$ is the gauge coset variable 
given on a link $<ij>$ by $g_{<ij>}\Gamma$, where $\Gamma \sim {\bf Z}_2$ is
the center of the gauge group $SU(2)$; $\sigma$ is a 2--form with 
values in $\{0, \pi \}\simeq {\bf Z}_2$.

The variables $U$ and $\sigma$ are not completely independent: it can be 
proved [42] that $e^{i\pi z_c (g)}$ defined in (5.9) is a function of the 
coset field $U$, which we denote by $e^{i\pi z_c(U)}$, and  
the following constraint holds:

$$
e^{i\pi z_c (U)}= e^{i\pi(d\sigma)_c} \eqno({\rm B}.1)
$$

for every cube $c$.

The partition function of model C can be rewritten as

$$
Z = \sum_\sigma \int \prod_{<ij>} d U_{<ij>} \prod_i d\Phi_i
$$
$$
e^{\sum_p |\chi| (U_{\partial p}) e^{i \sigma_p} - S_1(U, \Phi)}\prod_c 
\delta (e^{i[\pi z_c(U) -(d\sigma)_c]})
\eqno({\rm B}.2)
$$

where

$$
|\chi| (U_{\partial p}) = |\chi (g_{\partial p})|. \eqno({\rm B}.3)
$$

The introduction of the disorder field $D(\Sigma)$ induces a shift of 
$\sigma_p$ by $\Sigma_p$. We perform a duality transformation in the ${\bf 
Z}_2$ variable. Let $\tau$ denote the 1--form  
with values in $\{0,1\}$ dual to $\sigma$. Defining 

$$
S_0 (U; d\tau) = -\sum_p {\rm  ln} {\rm cosh}[|\chi| (U_{\partial p})]
+ ^* (d\tau)_p {\rm ln} {\rm tanh}[|\chi| (U_{\partial p})],
$$

the expectation value of 't Hooft disorder field can be written 

$$
\langle D(\Sigma) \rangle= {1\over Z} \sum_\tau \int \prod_{<ij>}d
U_{<ij>} 
\prod_i d\Phi_i e^{-[S_0(U; d\tau)+ S_1 (U,\Phi)]}
\prod_c e^{i[\pi z_c (U)+ (d\Sigma)_c |(^*\tau)_c}. \eqno({\rm B}.4)
$$

To exhibit the wordlines of magnetic currents we perform the abelian 
projection. We can decompose $U_{<ij>}$ as in the l.h.s. of (5.7), but with 
$\theta_{<ij>}$ restricted to the range $(-\pi, \pi)$.

As a consequence of $U(1)$--gauge invariance, we can expand 

$$
\int \prod_{<ij>} d C_{<ij>} e^{-[S_1 (C, \sigma_3)+ S_0 (C e^{i\sigma_3
{\theta\over 2}}; d\tau)]}  
\eqno({\rm B}.5)
$$

as a Fourier series in $d\theta$. The Fourier coefficients 
are denoted by $F(n; d\tau)$, where $n$ is an integer valued 2--form.

Defined the 1--form $\ell$ by

$$
\delta n = \ell \eqno({\rm B}.6)
$$

we decompose the 2--form $n$ as

$$
n= n [\ell] + ^*d\xi
$$

where $n[\ell]$ is an integer--valued solution of ({\rm B}.6) and $\xi$ a ${\bf 
Z}$--valued 1--form in the dual lattice. Furthermore, we define 
a ${\bf Z}/2$--valued 
1--form $\alpha$ in the dual lattice by

$$
\alpha =\xi + {1\over 2} \tau \eqno({\rm B}.7)
$$

and we adopt the notation:

$$
F (n [\ell] +{}^*d\xi ; d\tau)\equiv F (n[\ell] |
{}^*d\alpha)\eqno({\rm B}.8)
$$

Making use of eq.(5.10), we obtain 

$$
\langle D(\Sigma)\rangle = {1\over Z} \sum_{[\alpha]} 
\sum_{\ell:\delta\ell=0} F (n[\ell]|{}^*d\alpha)
$$
$$
\int \prod_{<ij>} d\theta_{<ij>} e^{i({}^*\alpha,2d \Sigma + 2\pi m(\theta))} 
e^{i(\theta, \ell)} \eqno({\rm B}.9)
$$

where $[\alpha]$ denotes a gauge equivalence class of $\alpha$. 

In eq.({\rm B}.9) we can replace $\alpha$ by a real--valued 1--form
$A$ by inserting the term
$\sum_{\rho:\delta\rho=0} e^{i(A,\rho)}$, 
where $\rho$ is a $4\pi {\bf Z}$--valued 1-- form.

We split $m(\theta)$ into a component of even magnetic charge $m_e(\theta)$
and a component of odd magnetic charge $m_o(\theta)$. Shifting $\rho$ 
by $2\pi m_e (\theta)$, we obtain the following identity:

$$
\langle D_\omega (j_x, j_y)\rangle = {1\over Z} \sum_{\ell:\delta\ell=0} 
\int d [A]
\sum_{\rho:\delta\rho=0} F(n [\ell] |{}^*dA) \int \prod_{<ij>} d\theta_{<ij>} 
e^{i(\theta,\ell)}
$$
$$
e^{i(A, 2 j_x + 2j_y + 2\omega + \rho + 2\pi m_o(\theta) + 4\pi 
\delta_{<R_+R_->})} \eqno({\rm B}.10)
$$

where $d[A]$ denotes formal integration over gauge equivalence classes of
$A$ . In eq.({\rm B}.10) worldlines of 't Hooft--Polyakov monopoles are 
described by $2 \omega + \rho$ and they exhibit sources at $\{x\}$ and 
$\{y\}$. Analogously $2\pi m_o$ describe the virtual 
trajectories of  ${\bf Z}_2$ monopoles. The representation ({\rm B}.10) 
shows also explicitely the independence of the choice of $\omega$ in the 
construction of Green functions of 't Hooft--Polyakov monopoles.

Taking $\omega$ with values in $\{0,\pi\}$, a similar representation
shows that
our construction of Green functions cannot be adapted to 
${\bf Z}_2$--monopoles, because a change in $\omega$ cannot be
reabsorbed by field redefinition.

\vskip 0.3truecm
{\bf Acknowledgements}
\vskip 0.3truecm

One of us (P.M.) gratefully acknowledges A. Di Giacomo and D. Marenduzzo 
for useful discussions and E.T.H. in Z\"urich for kind hospitality.

This work was partially supported by the European Commission TMR programme 
ERBFMPX--CT96--0045 to which P.M. is associated.

\vskip 0.5truecm
{\bf References}
\vskip 0.3truecm
\item{[1]} P.A.M. Dirac, Can. J. Phys. {\bf 33}, 650 (1955).

\item{[2]} J. Fr\"ohlich, P.A. Marchetti in ``The Algebraic Theory of 
Superselection Sectors. Introduction and Recent Results". D. Kastler ed., 
World Scientific 1990.

\item{[3]} J. Fr\"ohlich, P.A. Marchetti, Europhys. Lett. {\bf 2}, 933
(1986) and unpublished notes.

\item{[4]} N. Seiberg, E. Witten , Nucl. Phys. {\bf B426}, 19; {\bf B431}, 484  
(1994).

\item{[5]} G. 't Hooft, Nucl. Phys. {\bf B79}, 276 (1984); A.M. 
Polyakov JETP Lett. {\bf 20}, 194 (1974).

\item{[6]} B. Julia, A. Zee, Phys. Rev. {\bf D11}, 2227 (1975).

\item{[7]} A.M. Polyakov,Nucl. Phys.{\bf B120}; 429 (1977).

\item{[8]} H.A. Kramers, G.H. Wannier, Phys. Rev. {\bf 60}, 252 (1941);
F. Wegner, J. Math. Phys. {\bf 12}, 2259 (1971); G. 't Hooft, Nucl. 
Phys. {\bf B138}, 1 (1978).

\item{[9]} S. Mandelstam, Ann. Phys. {\bf 19}, 1 (1962). 

\item{[10]} G. 't Hooft, Nucl. Phys. {\bf B138}, 1 (1978); Nucl. Phys.
{\bf B153}, 141 (1979).

\item{[11]} J. L. Cardy, E. Rabinovici, Nucl. Phys. {\bf B205}, 1 
(1982); J.L. Cardy, Nucl. Phys. {\bf B205}, 17 (1982).

\item{[12]} J. Fr\"ohlich, T. Spencer, Commun Math. Phys. {\bf 83}, 
411 (1982).

\item{[13]} A. Guth, Phys. Rev. {\bf D21}, 2291 (1980).

\item{[14]} D. Brydges, E. Seiler, Journal Stat. Phys. {\bf 42}, 405 
(1986).

\item{[15]} E. Seiler, ``Gauge Theories as a Problem in Constructive 
Quantum Field Theory and Statistical Mechanics", Lecture Notes in Physics 
159, (1982).

\item{[16]} see e.g. C. Borgs, F. Nill, ``The Phase Diagram of Abelian 
Lattice Higgs Model; A Review of Rigorous Results". Preprint ETH 87--0069; 
J. Bricmont, J. Fr\"ohlich, Nucl. Phys. {\bf B230}, 407 (1984).

\item{[17]} J. Fr\"ohlich, P.A. Marchetti, Commun. Math. Phys. 
{\bf 121}, 177 (1989); V. M\"uller Z. Phys. {\bf C51}, 665 
(1991).

\item{[18]} R. Brower et al., Phys. Rev. {\bf D25}, 3319 (1982).

\item{[19]} J. Fr\"ohlich, P.A. Marchetti, Commun. Math. Phys. 
{\bf 112}, 343 (1987).

\item{[20]} J. Fr\"ohlich, P.A. Marchetti, Commun. Math. Phys. 
{\bf 116}, 127 (1988).

\item{[21]} J. Fr\"ohlich, P.A. Marchetti, Commun. Math. Phys. 
{\bf 121}, 177 (1989).

\item{[22]} E.C. Marino, B. Schroer, J.A. Swieca, Nucl. Phys. 
{\bf B200}, [FS 14], 472, (1982); K. Fredenhaegen, M. Marcu, Commun. 
Math. Phys. {\bf 92}, 81 (1983); K. Szlachanyi, Commun. Math. Phys. 
{\bf 108}, 319 (1987); J. Fr\"ohlich, P.A. Marchetti, Lett. Math. 
Phys. {\bf 16}, 347 (1988); Nucl. Phys. B355, 1 (1990); P.A. 
Marchetti, Europhys. Lett. {\bf 4}, 633 (1987).

\item{[23]} J. Bricmont, J. Fr\"ohlich, Nucl. Phys. {\bf B251}, [FS 
13], 517 (1985); Commun. Math. Phys. {\bf 98}, 553 (1985); P.A. 
Marchetti, Commun. Math. Phys. {\bf 117}, 501 (1988).

\item{[24]} J. Fr\"ohlich in ``Scaling and Self--similarity in Physics", J.
Fr\"ohlich ed., ``Progress in Physics", Birkhauser, 1983.

\item{[25]} T. Kennedy, C.King, Phys. Rev. Lett. {\bf 55}, 776 (1985);
Commun. Math. Phys. {\bf 104}, 345 (1986).

\item{[26]} D. Buchholz, Commun. Math. Phys. {\bf 85}, 49 (1982).

\item{[27]} L. Polley, U.J. Wiese, Nucl. Phys.{\bf B356}, 629 (1991);
M.I. Polikarpov, L. Polley, U.J. Wiese, Phys. Lett. {\bf B253}, 212
(1991); L. Del Debbio, A. di Giacomo, G. Paffuti, Phys. Lett. 
{\bf B349}, 513 (1995).

\item{[28]} S. Coleman in ``The Unity of the Fundamental Interactions"
(Erice 1981), A. Zichichi ed., Plenum Press, 1983.

\item{[29]} D. Durhuus, J. Fr\"ohlich, Commun. Math. Phys. {\bf 75}, 
103 (1988).

\item{[30]} D. Marenduzzo, Tesi di laurea, University of Padova. 

\item{[31]} J. Fr\"ohlich in ``Unified Theories of Elementary Particles. 
Critical Assessments and Prospects". P. Breitenhohner, H.P. Durr eds.,   
Springer Lecture Notes in Physics vol. 160 (1982).

\item{[32]} D. Brydges, J. Fr\"ohlich, T. Spencer, Commun. Math. Phys. 
{\bf 83}, 123 (1982).

\item{[33]} L. Del Debbio, A. Di Giacomo, G. Paffutti, L. Pieri, Phys. 
Lett. {\bf B355}, 255 (1995); A. Di Giacomo, G. Paffuti, Phys. Rev. D 
{\bf 56}, 6816 (1997); A. I. Veselov, M.I. Polikarpov, M.N. Chernodub,
JETP Letters {\bf 63}, 411 (1996).

\item{[34]} E. Witten, Phys. Lett. {\bf B86}, 283 (1979).

\item{[35]} A. Jaffe, C. Taubes ``Vortices and Monopoles. Structure of 
a static gauge theories". Progress in Phys. {\bf 2}; A. Jaffe, D. Ruelle
eds., Birkhauser 1980.

\item{[36]} G. 't Hooft, Nucl. Phys. {\bf B120}, 455 (1981).

\item{[37]} F. Englert in ``Hadron Structure and Lepton--Hadron 
Interactions" (Cargese 1977) M. Levy et al. eds., Plenum Press, 1979.

\item{[38]} A.S. Kronfeld, G. Schierholz, U.J. Wiese, Nucl. Phys. 
{\bf B293}, 461 (1987).

\item{[39]} A.S. Kronfeld, M.L. Laursen, G. Schierholz, U.J. Wiese, Phys. 
Lett. {\bf B198}, 516 (1987).

\item{[40]} A.S. Cattaneo et al., ``Four--Dimensional Yang--Mills theory as
a Deformation of Topological BF theory";
hep--th/9705123, to appear in Commun. Math. Phys.

\item{[41]} B. Julia, A. Zee, Phys. Rev. {\bf D11}, 2227 (1975).

\item{[42]} G. Mack, V.B. Petkova, Z. Phys. {\bf C12}, 177 (1982).
\vfill\eject
\epsfysize=19.5cm\epsfbox{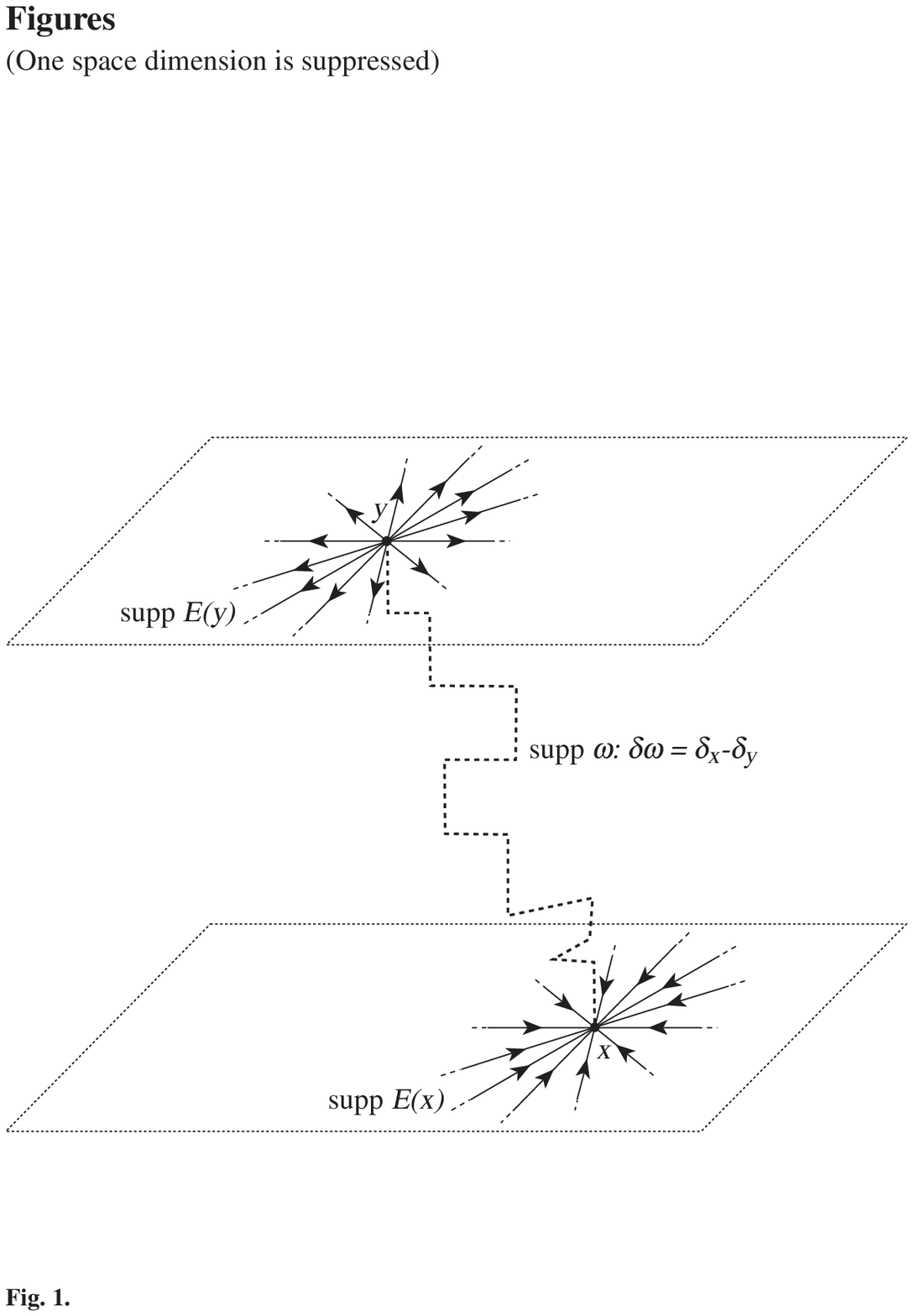}
\vfill\eject
\epsfysize=19.5cm\epsfbox{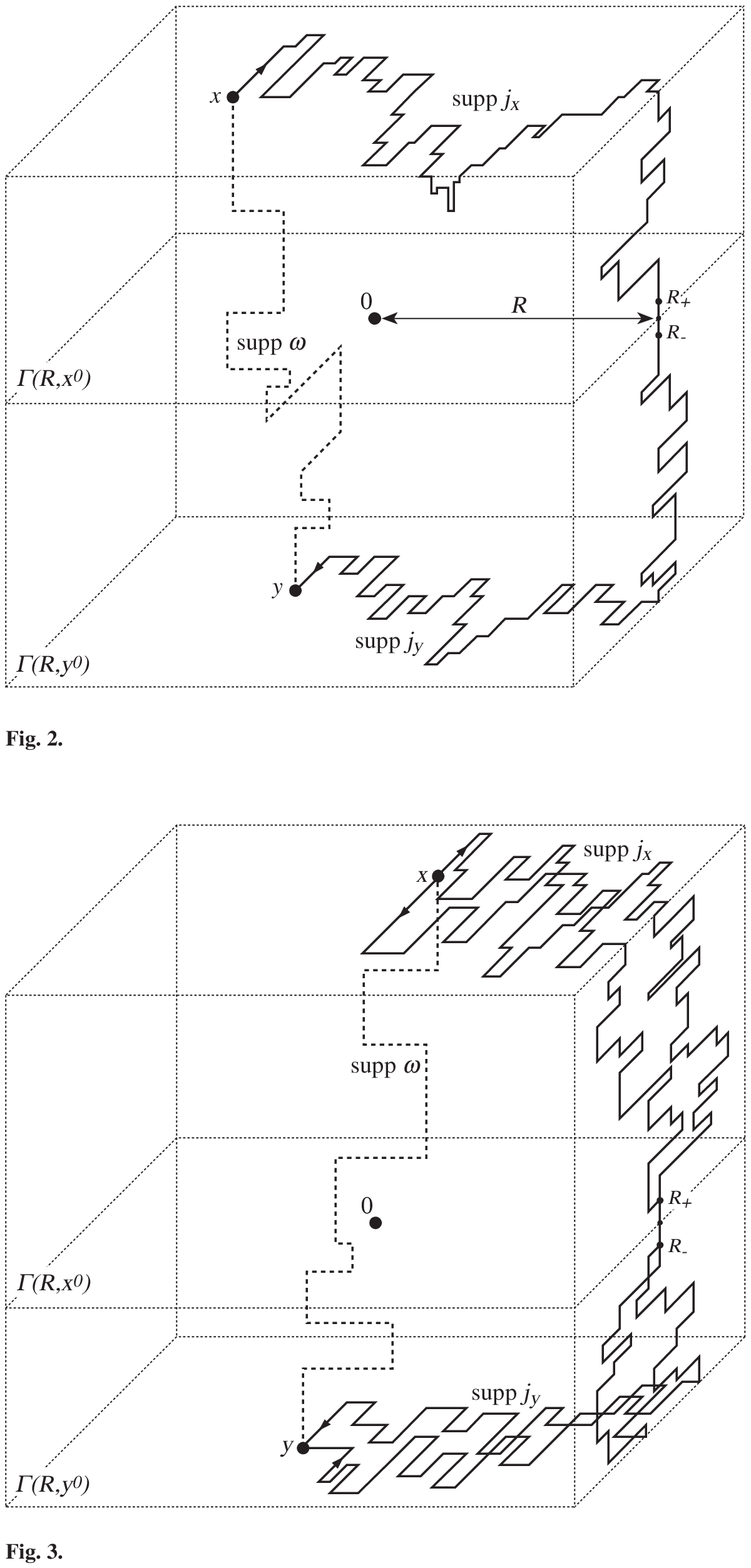}

\bye